\documentclass{jpsj3}
\usepackage{txfonts}
\usepackage{braket}
\usepackage{color}

\title{Photoinduced Pseudospin Polarization in a Three-Orbital Hubbard Model}

\author{Kenji Yonemitsu\thanks{E-mail: kxy@phys.chuo-u.ac.jp}}
\inst{Department of Physics, Chuo University, Bunkyo, Tokyo 112-8551, Japan 
} 

\abst{ In a Hubbard model for the Kitaev spin-liquid candidate material $\alpha$-RuCl$_3$ with three $t_{2g}$ orbitals per Ru site, we calculate photoinduced dynamics based on the exact diagonalization method and interpret them with the help of a high-frequency expansion in quantum Floquet theory. The high-frequency expansion shows two types of effective magnetic fields during the application of a circularly polarized light field. One of them originates from spin-orbit coupling and is within the honeycomb lattice. The other is of purely kinetic origin and perpendicular to the lattice. The former fields are antiparallel at the two sites within a unit cell and rotate in accordance with the momentum distribution of holes that follow the light field. When the light field is weak, pseudospin dynamics are governed by the former fields; thus, the average of the pseudospins almost vanishes. The latter fields are parallel at the two sites within a unit cell and produce nonzero perpendicular components of the pseudospins when the light field is strong. Numerically obtained perpendicular components are consistent with the latter fields when the frequency of the light field is well below the Mott gap. The relevance to the inverse Faraday effect recently observed in $\alpha$-RuCl$_3$ is discussed. }

\begin{document}
\maketitle

\section{Introduction}
Photoinduced cooperative phenomena, including photoinduced phase transitions, have attracted considerable attention.\cite{basov_rmp11,giannetti_aip16,nicoletti_aop16,chergui_cr17,kaiser_ps17,basov_nmat17,kawakami_jpb18,oka_arcmp19,ishihara_jpsj19,delatorre_rmp21,koshihara_pr22} For those accompanied by a symmetry change, its mechanisms have been a central issue. If the transition is between metastable and stable phases, photoexcitations produce seeds of fluctuations that grow under some conditions.\cite{nagaosa_prb89,miyano_prl97,fiebig_s98} However, recent experimental techniques have realized the photoinduced coherent motion of many electrons.\cite{kawakami_prl10,matsubara_prb14} On a timescale shorter than the electron-electron scattering time, inversion symmetry can be broken to cause second harmonic generation in centrosymmetric systems.\cite{pronin_prb94,silva_np18,ikeda_pra18,kawakami_ncomms20} 

As to changes in the time-reversal symmetry, there are photoinduced phase transitions between the ferromagnetic metal and the antiferromagnetic insulator in perovskite manganites,\cite{miyano_prl97,fiebig_s98,ishihara_jpsj19} which are often treated in the double exchange model. These transitions are believed to be achieved by domain growth from the seeds of fluctuations. For intended symmetry breaking of the time-reversal symmetry, Floquet topological insulators with peculiar Hall responses\cite{oka_prb09,kitagawa_prb11,ezawa_prl13,schueler_prx20}, spin chirality or current generation,\cite{sato_prl16} chiral spin liquids,\cite{claassen_ncomms17,kitamura_prb17} topological superconductivity,\cite{takasan_prb17,chono_prb20} and spin polarization\cite{mochizuki_apl18} are theoretically proposed to be induced by circularly polarized light fields, as well as magnetization by rotating magnetic fields.\cite{takayoshi_prb14a,takayoshi_prb14b} Recently, the inverse Faraday effect has been observed during the application of a circularly polarized light field to $\alpha$-RuCl$_3$ in the quantum-spin-liquid phase.\cite{amano_prr22} It suggests photoinduced polarization of magnetic moments that is perpendicular to the honeycomb lattice. In a different context, the inverse Faraday effect on tight-binding models with the Rashba-type spin-orbit coupling or the pseudospin inversion asymmetry spin-orbit coupling is theoretically studied using the time-dependent Schr\"odinger equation as well as Floquet theory.\cite{tanaka_njp20,okyay_prb20} Aside from current-induced spin polarization,\cite{j_q_li_prb21} photoinduced magnetization is theoretically discussed from the viewpoint of the nonlinear Edelstein effect.\cite{h_xu_prb21} 

In quantum-spin-liquid phases, magnetic moments are completely suppressed by frustrations. Photoexcitations may bring about nonzero magnetic moments in effect by reducing frustrations. Above all, $ \alpha $-RuCl$_3$ has attracted much attention as a candidate material that possesses Majorana fermions.\cite{takagi_nrevphys19,motome_jpsj20}. It is a Mott insulator, and its pseudospin degrees of freedom in the quantum-spin-liquid phase are now believed to be basically described by the Kitaev model\cite{kitaev_ap06,jackeli_prl09} and Majorana fermions\cite{nasu_np16,kasahara_n18} on the two-dimensional honeycomb lattice. Nonequilibrium properties of Kitaev quantum spin liquids are also theoretically investigated.\cite{nasu_prr19,minakawa_prl20,arakawa_prb21,kanega_prr21,arakawa_prb21full}

In this work, we theoretically study photoinduced dynamics of pseudospins and their relevance to experimental observations in $ \alpha $-RuCl$_3$.\cite{amano_prr22} Since near- and/or mid-infrared pulses are used, the charge degrees of freedom are also important. Thus, we consider a Hubbard model consisting of $d_{yz}$, $d_{xz}$, and $d_{xy}$ orbitals on each site of the honeycomb lattice.\cite{rau_prl14,h_s_kim_prb16,winter_prb16,eichstaedt_prb19} We use the model parameters employed in Ref.~\citen{winter_prb16}, but the hopping term is limited to the nearest neighbors because of the smallness of the system treated by the exact diagonalization method. 

Intersite Coulomb interactions considered and evaluated in Ref.~\citen{eichstaedt_prb19} are found to cause strong excitonic effects, which are inconsistent with reported optical conductivity spectra\cite{sandilands_prb16a,sandilands_prb16b,reschke_jpcm18,warzanowski_prr20} and electron energy loss spectroscopy measurements.\cite{koitzsch_prl16} Therefore, we do not take them into account. From the observed spectra,\cite{sandilands_prb16a,sandilands_prb16b,koitzsch_prl16,reschke_jpcm18,warzanowski_prr20} $ \alpha $-RuCl$_3$ is a Mott insulator with a gap of about 1 eV. Below the Mott gap, in-gap states are observed, the assignment of which is still controversial. In this study, they are qualitatively reproduced by the exact diagonalization method and regarded basically as spin-orbit excitons.\cite{b_h_kim_prl16,warzanowski_prr20,lee_npjqm21} 

We consider only the quantum-spin-liquid phase. To take account of such strong electron correlations in the ground state, we use the exact diagonalization method. Because the $z$-component of the total spin is not conserved owing to spin-orbit coupling, we simultaneously treat six spin orbitals per site. Numerically, we consider a minimum-size system that has three unit cells, each of which consists of two sites A and B, and use the periodic boundary conditions shown in Fig.~\ref{fig:honeycomb} to maintain the threefold symmetry. Note that the employed model parameters (transfer integrals and crystal fields)\cite{winter_prb16} do not have the threefold symmetry. By using this system, we do not consider the antiferromagnetic state of collinear zigzag type,\cite{winter_prb16} which is the experimentally observed ground state and is not threefold-symmetric. As a consequence of the smallness of the system, the excitation spectra are quite discrete, and the low-energy pseudospin degrees of freedom, which are often described by Majorana fermions, may not be quantitatively treated. On the other hand, we can directly treat both $J_{\mathrm{eff}}=\frac12$ and $J_{\mathrm{eff}}=\frac32$ pseudospins and the corresponding charge degrees of freedom to cover a large energy scale ranging from on-site Coulomb interactions to excitations below the Mott gap. 

The pseudospin and other dynamics are calculated on the basis of numerical solutions to the time-dependent Schr\"odinger equation. They are interpreted with the help of a high-frequency expansion in quantum Floquet theory, which gives effective Hamiltonian terms during a continuous-wave excitation before thermalization.\cite{rahav_pra03,mananga_jcp11,goldman_prx14,eckardt_njp15,itin_prl15,bukov_ap15,mikami_prb16,yonemitsu_jpsj17b} The effective term originating from spin-orbit coupling shows that photoexcitations produce effective in-plane magnetic fields acting on pseudospins. The emergence and initial dynamics of pseudospins induced by a circularly polarized light field are numerically consistent with these effective magnetic fields. The commutators among the kinetic operators on the three bonds produce effective out-of-plane magnetic fields that polarize pseudospins in the direction perpendicular to the honeycomb lattice, i.e., they give rise to the inverse Faraday effect. We will discuss the assignment of in-gap states below the Mott gap and how pseudospins are polarized during the application of a circularly polarized light field. 

\section{Three-Orbital Model on Honeycomb Lattice}
\begin{figure}
\includegraphics[height=10cm]{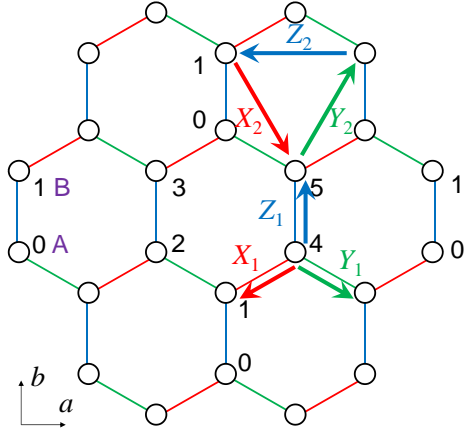}
\caption{(Color online) 
Honeycomb lattice with periodic boundary conditions as indicated by numbers. Vectors {\boldmath $X_1$}, {\boldmath $Y_1$}, and {\boldmath $Z_1$} connect nearest-neighbor sites, and {\boldmath $X_2$}, {\boldmath $Y_2$}, and {\boldmath $Z_2$} connect next-nearest-neighbor sites, as shown here. 
\label{fig:honeycomb}}
\end{figure}

We employ the hole picture and treat the case of one hole per site consisting of three $t_{2g}$ orbitals. Following Ref.~\citen{winter_prb16}, we use a three-orbital Hubbard model, 
\begin{equation}
H = H_{\mathrm{hop}} + H_{\mathrm{CF}} + H_{\mathrm{SO}} + H_U
\;, \label{eq:model}
\end{equation}
which consists of the kinetic term, the crystal-field term, spin-orbit coupling, and Coulomb interactions, respectively. In what follows, we use 
\begin{equation}
\vec{\mathbf{c}}_i^\dagger = \left( 
c_{i,yz,\uparrow}^\dagger \; c_{i,yz,\downarrow}^\dagger \; 
c_{i,xz,\uparrow}^\dagger \; c_{i,xz,\downarrow}^\dagger \; 
c_{i,xy,\uparrow}^\dagger \; c_{i,xy,\downarrow}^\dagger 
\right)
\;, \label{eq:creation}
\end{equation}
where $ c_{i,a,\sigma}^\dagger $ creates a hole in orbital $ a \in \{ yz, xz, xy \} $ with spin $ \sigma $ at site $ i $. The kinetic term is described by 
\begin{equation}
H_{\mathrm{hop}} = -\sum_{ij} \vec{\mathbf{c}}_i^\dagger \left\{ \mathbf{T}_{ij} \otimes \mathbb{I}_{2\times2} \right\} \vec{\mathbf{c}}_j
\;, \label{eq:kinetic}
\end{equation}
where $ \mathbb{I}_{2\times2} $ is the 2$\times$2 identity matrix and $ \mathbf{T}_{ij} $ is the hopping matrix defined for each bond connecting nearest-neighbor sites $ i $ and $ j $. The latter is one of $ \mathbf{T}_1^{\mathrm{X}} $, $ \mathbf{T}_1^{\mathrm{Y}} $, and $ \mathbf{T}_1^{\mathrm{Z}} $ for the X$_1$, Y$_1$, and Z$_1$ bonds, respectively, shown in Fig.~\ref{fig:honeycomb}. 
They are given by 
\begin{equation}
\mathbf{T}_1^{\mathrm{X}} = 
\begin{pmatrix}
t^{\,\prime}_{3 } & t^{\,\prime}_{4a} & t^{\,\prime}_{4b} \\
t^{\,\prime}_{4a} & t^{\,\prime}_{1a} & t^{\,\prime}_{2 } \\
t^{\,\prime}_{4b} & t^{\,\prime}_{2 } & t^{\,\prime}_{1b} \\
\end{pmatrix}
\,, 
\mathbf{T}_1^{\mathrm{Y}} = 
\begin{pmatrix}
t^{\,\prime}_{1a} & t^{\,\prime}_{4a} & t^{\,\prime}_{2 } \\
t^{\,\prime}_{4a} & t^{\,\prime}_{3 } & t^{\,\prime}_{4b} \\
t^{\,\prime}_{2 } & t^{\,\prime}_{4b} & t^{\,\prime}_{1b} \\
\end{pmatrix}
\;, \label{eq:T1XY}
\end{equation}
and 
\begin{equation}
\mathbf{T}_1^{\mathrm{Z}} = 
\begin{pmatrix}
t_{1} & t_{2} & t_{4} \\
t_{2} & t_{1} & t_{4} \\
t_{4} & t_{4} & t_{3} \\
\end{pmatrix}
\; \label{eq:T1Z}
\end{equation}
with $ t_{1 } $=0.0509 eV, 
$ t^{\,\prime}_{1a} $=0.0449 eV, 
$ t^{\,\prime}_{1b} $=0.0458 eV, 
$ t_{2 } $=0.1582 eV, 
$ t^{\,\prime}_{2 } $=0.1622 eV, 
$ t_{3 } $=$-$0.1540 eV, 
$ t^{\,\prime}_{3 } $=$-$0.1031 eV, 
$ t_{4 } $=$-$0.0202 eV, 
$ t^{\,\prime}_{4a} $=$-$0.0151 eV, and 
$ t^{\,\prime}_{4b} $=$-$0.0109 eV. 
The crystal-field term is described by 
\begin{equation}
H_{\mathrm{CF}} = -\sum_{i} \vec{\mathbf{c}}_i^\dagger \left\{ \mathbf{E}_{i} \otimes \mathbb{I}_{2\times2} \right\} \vec{\mathbf{c}}_i
\;, \label{eq:crystal_field}
\end{equation}
where $ \mathbf{E}_{i} $ is the crystal-field tensor given by 
\begin{equation}
\mathbf{E}_{i} = 
\begin{pmatrix}
0          & \Delta_{1} & \Delta_{2} \\
\Delta_{1} & 0          & \Delta_{2} \\
\Delta_{2} & \Delta_{2} & \Delta_{3} \\
\end{pmatrix}
\; \label{eq:E_i}
\end{equation}
with $ \Delta_{1} $=$-$0.0198 eV, $ \Delta_{2} $=$-$0.0175 eV, and $ \Delta_{3} $=$-$0.0125 eV. The spin-orbit coupling is given by 
\begin{equation}
H_{\mathrm{SO}} = \frac{\lambda}{2}\sum_{i} \vec{\mathbf{c}}_i^\dagger 
\begin{pmatrix}
0 & -i\sigma_z & i\sigma_y \\
i\sigma_z & 0 & -i\sigma_x \\
-i\sigma_y & i\sigma_x & 0 \\
\end{pmatrix}
\vec{\mathbf{c}}_i
\;, \label{eq:spin_orbit}
\end{equation}
where $ \sigma_x $, $ \sigma_y $, and $ \sigma_z $ are the Pauli matrices and $ \lambda $=0.15 eV. The Coulomb interactions are given by 
\begin{eqnarray}
H_U & = & U \sum_{i,a} n_{i,a,\uparrow} n_{i,a,\downarrow}
+( U^{\,\prime} - J_{\mathrm{H}} ) 
\sum_{i,a<b,\sigma} n_{i,a,\sigma} n_{i,b,\sigma} \nonumber \\
& & + U^{\,\prime} \sum_{i,a \neq b} n_{i,a,\uparrow} n_{i,b,\downarrow}
- J_{\mathrm{H}} \sum_{i,a \neq b} 
c^\dagger_{i,a,\uparrow} c_{i,a,\downarrow} 
c^\dagger_{i,b,\downarrow} c_{i,b,\uparrow} \nonumber \\
& & + J_{\mathrm{H}} \sum_{i,a \neq b} 
c^\dagger_{i,a,\uparrow} c^\dagger_{i,a,\downarrow} 
c_{i,b,\downarrow} c_{i,b,\uparrow}
\;, \label{eq:interaction}
\end{eqnarray}
where $ n_{i,a,\sigma}=c^\dagger_{i,a,\sigma} c_{i,a,\sigma} $, $ U $ is the intraorbital Coulomb repulsion, $ J_{\mathrm{H}} $ is Hund's coupling strength, and $ U^{\,\prime} = U - 2 J_{\mathrm{H}} $ is the interorbital repulsion with $ U $=3.0 eV and $ J_{\mathrm{H}} $=0.6 eV. 

To describe $J_{\mathrm{eff}}=\frac12$ states, we introduce 
\begin{equation}
p_{i,\uparrow}^\dagger = \frac{1}{\sqrt{3}}\left( 
- c^\dagger_{i,xy,\uparrow}
-ic^\dagger_{i,xz,\downarrow}
- c^\dagger_{i,yz,\downarrow}
\right) 
\;, \label{eq:p_up}
\end{equation}
\begin{equation}
p_{i,\downarrow}^\dagger = \frac{1}{\sqrt{3}}\left( 
  c^\dagger_{i,xy,\downarrow}
+ic^\dagger_{i,xz,\uparrow}
- c^\dagger_{i,yz,\uparrow}
\right) 
\;, \label{eq:p_down}
\end{equation}
which give $\Ket{\frac12,\frac12}_i$=$p_{i,\uparrow}^\dagger \Ket{0} $ and 
$ \Ket{\frac12,-\frac12}_i$=$p_{i,\downarrow}^\dagger \Ket{0} $ at site $ i $ (their eigenvalue of $ H_{\mathrm{SO}} $ is $ -\lambda $).
To describe $J_{\mathrm{eff}}=\frac32$ states, we introduce 
\begin{equation}
q_{i,3/2}^\dagger = \frac{1}{\sqrt{2}}\left( 
-ic^\dagger_{i,xz,\uparrow}
- c^\dagger_{i,yz,\uparrow}
\right) 
\;, \label{eq:q_p3}
\end{equation}
\begin{equation}
q_{i,1/2}^\dagger = \frac{1}{\sqrt{6}}\left( 
2 c^\dagger_{i,xy,\uparrow}
-ic^\dagger_{i,xz,\downarrow}
- c^\dagger_{i,yz,\downarrow}
\right) 
\;, \label{eq:q_p1}
\end{equation}
\begin{equation}
q_{i,-1/2}^\dagger = \frac{1}{\sqrt{6}}\left( 
2 c^\dagger_{i,xy,\downarrow}
-ic^\dagger_{i,xz,\uparrow}
+ c^\dagger_{i,yz,\uparrow}
\right) 
\;, \label{eq:q_m1}
\end{equation}
\begin{equation}
q_{i,-3/2}^\dagger = \frac{1}{\sqrt{2}}\left( 
-ic^\dagger_{i,xz,\downarrow}
+ c^\dagger_{i,yz,\downarrow}
\right) 
\;, \label{eq:q_m3}
\end{equation}
giving $\Ket{\frac32,\frac32}_i$=$q_{i,3/2}^\dagger \Ket{0} $, 
$ \Ket{\frac32,\frac12}_i$=$q_{i,1/2}^\dagger \Ket{0} $, 
$ \Ket{\frac32,-\frac12}_i$=$q_{i,-1/2}^\dagger \Ket{0} $, and 
$ \Ket{\frac32,-\frac32}_i$=$q_{i,-3/2}^\dagger \Ket{0} $ at site $ i $ (their eigenvalue of $ H_{\mathrm{SO}} $ is $ \lambda/2 $). For $J_{\mathrm{eff}}=\frac12$, the pseudospin densities are thus described by 
\begin{equation}
j^{(1/2)}_{i,x} = \frac12 \Braket{ p_{i,\uparrow}^\dagger p_{i,\downarrow} 
+ p_{i,\downarrow}^\dagger p_{i,\uparrow} }
\;, \label{eq:j12x}
\end{equation}
\begin{equation}
j^{(1/2)}_{i,y} = \frac12 \Braket{ -i p_{i,\uparrow}^\dagger p_{i,\downarrow} 
+i p_{i,\downarrow}^\dagger p_{i,\uparrow} }
\;, \label{eq:j12y}
\end{equation}
\begin{equation}
j^{(1/2)}_{i,z} = \frac12 \Braket{ p_{i,\uparrow}^\dagger p_{i,\uparrow} 
- p_{i,\downarrow}^\dagger p_{i,\downarrow} }
\;, \label{eq:j12z}
\end{equation}
which give 
\begin{equation}
j^{(1/2)}_{i,\perp} = \frac{1}{\sqrt{3}}\left( 
j^{(1/2)}_{i,x} + j^{(1/2)}_{i,y} + j^{(1/2)}_{i,z}
\right) 
\;, \label{eq:j12perp}
\end{equation}
as the component perpendicular to the honeycomb lattice. The corresponding charge density is written as 
\begin{equation}
\rho^{(1/2)}_{i} = \Braket{ p_{i,\uparrow}^\dagger p_{i,\uparrow} 
+ p_{i,\downarrow}^\dagger p_{i,\downarrow} }
\;. \label{eq:rho12}
\end{equation}
For $J_{\mathrm{eff}}=\frac32$, they are described with 
\begin{equation}
\vec{\mathbf{q}}_i^\dagger = \left( 
q_{i, 3/2}^\dagger \; q_{i, 1/2}^\dagger \; 
q_{i,-1.2}^\dagger \; q_{i,-3/2}^\dagger 
\right)
\;  \label{eq:q_dagger}
\end{equation}
by 
\begin{equation}
j^{(3/2)}_{i,x} = \braket{\vec{\mathbf{q}}_i^\dagger 
\begin{pmatrix}
0          & \sqrt{3}/2 & 0          & 0          \\
\sqrt{3}/2 & 0          & 1          & 0          \\
0          & 1          & 0          & \sqrt{3}/2 \\
0          & 0          & \sqrt{3}/2 & 0          \\
\end{pmatrix}
\vec{\mathbf{q}}_i}
\;, \label{eq:j32x}
\end{equation}
\begin{equation}
j^{(3/2)}_{i,y} = \braket{\vec{\mathbf{q}}_i^\dagger 
\begin{pmatrix}
0           & -i\sqrt{3}/2 & 0           & 0            \\
i\sqrt{3}/2 & 0            & -i          & 0            \\
0           & i            & 0           & -i\sqrt{3}/2 \\
0           & 0            & i\sqrt{3}/2 & 0            \\
\end{pmatrix}
\vec{\mathbf{q}}_i}
\;, \label{eq:j32y}
\end{equation}
\begin{equation}
j^{(3/2)}_{i,z} = \braket{\vec{\mathbf{q}}_i^\dagger 
\begin{pmatrix}
3/2 & 0   & 0    & 0    \\
0   & 1/2 & 0    & 0    \\
0   & 0   & -1/2 & 0    \\
0   & 0   & 0    & -3/2 \\
\end{pmatrix}
\vec{\mathbf{q}}_i}
\;, \label{eq:j32z}
\end{equation}
which give 
\begin{equation}
j^{(3/2)}_{i,\perp} = \frac{1}{\sqrt{3}}\left( 
j^{(3/2)}_{i,x} + j^{(3/2)}_{i,y} + j^{(3/2)}_{i,z}
\right) 
\;, \label{eq:j32perp}
\end{equation}
as the component perpendicular to the honeycomb lattice. Note that the magnetic moment (in units of Bohr magneton $\mu_{\mathrm{B}}$) $\mbox{\boldmath $M$}=-\mbox{\boldmath $l$}+2\mbox{\boldmath $s$}$\cite{jackeli_prl09,takayama_jpsj21} with an effective angular momentum $\mbox{\boldmath $l$}$ and a hole spin operator $\mbox{\boldmath $s$}$ is given by $ \mbox{\boldmath $M$}=-2 \mbox{\boldmath $j$}^{(1/2)} $ (i.e., the $g$ factor is $-2$ when the covalency factor is omitted)\cite{rau_annrev16,takagi_nrevphys19,motome_jpsj20,takayama_jpsj21} at each site. 

The initial state is the ground state obtained by the exact diagonalization method applied to the six-site system shown in Fig.~\ref{fig:honeycomb}. Photoexcitation is introduced through the Peierls phase 
\begin{equation}
c_{i,a,\sigma}^\dagger c_{j,b,\sigma} \rightarrow
\exp \left[
-\frac{ie}{\hbar c} \mbox{\boldmath $r$}_{ij} \cdot \mbox{\boldmath $A$}(t)
\right] c_{i,a,\sigma}^\dagger c_{j,b,\sigma}
\;, \label{eq:photo_excitation}
\end{equation}
which is substituted into Eq.~(\ref{eq:kinetic}) for each combination of sites $ i $ and $ j $ with relative position $ \mbox{\boldmath $r$}_{ij}=\mbox{\boldmath $r$}_j-\mbox{\boldmath $r$}_i $. When parameter values are referred to, we use $e$=$a$=$\hbar$=1 with $ a $ being the intersite distance. The optical conductivity spectra are calculated for the ground state as before.\cite{yonemitsu_jpsj11b} For photoinduced dynamics, we employ symmetric $n$-cycle electric-field pulses with $n$ being an integer of duration $ t_{\mathrm{off}}=2n\pi/\omega $ with frequency $\omega$. Thus, the time-dependent vector potential for a circularly polarized light field is written as 
\begin{equation}
A_{a}(t) = \theta(0<t<t_{\mathrm{off}})
\frac{c(-F_{\mathrm{L}}-F_{\mathrm{R}})}{\omega}
\left[ \sin(\omega t-\theta_\text{ini})-\sin(-\theta_\text{ini}) \right]
\;, \label{eq:a_circular_a}
\end{equation}
\begin{equation}
A_{b}(t) = \theta(0<t<t_{\mathrm{off}})
\frac{c(F_{\mathrm{L}}-F_{\mathrm{R}})}{\omega}
\left[ \cos(\omega t-\theta_\text{ini})-\cos(-\theta_\text{ini}) \right]
\;, \label{eq:a_circular_b}
\end{equation}
with $\theta(0<t<t_{\mathrm{off}})$=1 for $0<t<t_{\mathrm{off}}$ and $\theta(0<t<t_{\mathrm{off}})$=0 otherwise for the $a$ and $b$ (not to be confused with orbital) components within the plane (Fig.~\ref{fig:honeycomb}). It corresponds to the electric field given by 
\begin{equation}
E_{a}(t) = \theta(0<t<t_{\mathrm{off}})
(F_{\mathrm{L}}+F_{\mathrm{R}}) \cos(\omega t-\theta_\text{ini})
\;, \label{eq:e_circular_a}
\end{equation}
\begin{equation}
E_{b}(t) = \theta(0<t<t_{\mathrm{off}})
(F_{\mathrm{L}}-F_{\mathrm{R}}) \sin(\omega t-\theta_\text{ini})
\;. \label{eq:e_circular_b}
\end{equation}
Throughout this paper, we set $F_{\mathrm{R}}$=0 when $F_{\mathrm{L}} \neq$0 and $F_{\mathrm{L}}$=0 when $F_{\mathrm{R}} \neq$0. Considering that the distance between neighboring Ru$^{+3}$ ions is 3.45 \AA,\cite{cao_prb16} $F_{\mathrm{L}}$=0.02 ($F_{\mathrm{L}}$=0.2) used later corresponds to 0.58 MV/cm (5.8 MV/cm). 
The time-dependent Schr\"odinger equation is numerically solved by expanding the exponential evolution operator with a time slice $ dt $=0.02 to the 15th order and by checking the conservation of the norm.\cite{yonemitsu_prb09}

\section{Consequence of Reflection Symmetry}
The two-dimensional system considered here has reflection symmetry with respect to a line ($\parallel b$-axis) containing a Z$_1$ bond and a line ($\parallel a$-axis) containing the perpendicular bisector of a Z$_1$ bond. Photoinduced dynamics conform to the corresponding symmetry operations (within the two-dimensional system), as explained below. 

\subsection{Reflection symmetry with respect to Z$_1$ \label{sect:reflection}}
For $ \mbox{\boldmath $A$}(t) $=0, this reflection symmetry makes the model [Eq.~(\ref{eq:model})] invariant under the operation consisting of the exchange of the X$_1$ and Y$_1$ bonds, that of the $d_{xz}$ and $d_{yz}$ orbitals, $c_{i,a,\downarrow} \rightarrow -i c_{i,a,\downarrow}$ ($c_{i,a,\downarrow}^\dagger \rightarrow i c_{i,a,\downarrow}^\dagger$), and then the exchange of the $\sigma=\uparrow$ and $\sigma=\downarrow$ operators. This symmetry operation exchanges a light field of left-hand circular polarization with $ F_{\mathrm{L}} $ and $ \theta_\text{ini}=\theta_0 $ and a light field of right-hand circular polarization with $ F_{\mathrm{R}} $ of the same magnitude and $ \theta_\text{ini}=\theta_0+\pi $. It also exchanges $ j^{( J_{\mathrm{eff}} )}_{i,x}  $ for $ - j^{( J_{\mathrm{eff}} )}_{i,y}  $ and $ j^{( J_{\mathrm{eff}} )}_{i,z}  $ for $ -j^{( J_{\mathrm{eff}} )}_{i,z}  $. As a consequence, if a light field with $ F_{\mathrm{L}} $ and $ \theta_\text{ini}=\theta_0 $ produces $ j^{( J_{\mathrm{eff}} )}_{i,\perp} $ at time $ t $, it guarantees that a light field with $ F_{\mathrm{R}} $ of the same magnitude and $ \theta_\text{ini}=\theta_0+\pi $ produces $ - j^{( J_{\mathrm{eff}} )}_{i,\perp} $ (i.e., of the same magnitude and opposite sign) at the same time $ t $, which is numerically confirmed. 

\subsection{Reflection symmetry with respect to $\perp$ bisector of Z$_1$ \label{sect:refl_bisector}}
For $ \mbox{\boldmath $A$}(t) $=0, this reflection symmetry makes the model [Eq.~(\ref{eq:model})] invariant under the operation consisting of the exchange of sites A and B connected through the Z$_1$ bond, that of the X$_1$ and Y$_1$ bonds, that of the $d_{xz}$ orbital at site A (B) and the $d_{yz}$ orbital at site B (A), $c_{i,a,\downarrow} \rightarrow -i c_{i,a,\downarrow}$ ($c_{i,a,\downarrow}^\dagger \rightarrow i c_{i,a,\downarrow}^\dagger$), and then the exchange of the $\sigma=\uparrow$ and $\sigma=\downarrow$ operators. This symmetry operation exchanges a light field of left-hand circular polarization with $ F_{\mathrm{L}} $ and $ \theta_\text{ini}=\theta_0 $ and a light field of right-hand circular polarization with $ F_{\mathrm{R}} $ of the same magnitude and $ \theta_\text{ini}=\theta_0 $. It also exchanges $ j^{( J_{\mathrm{eff}} )}_{{\mathrm{site A}},x}  $ for $ -j^{( J_{\mathrm{eff}} )}_{ {\mathrm{site B}},y}  $, $ j^{( J_{\mathrm{eff}} )}_{{\mathrm{site A}},y}  $ for $ -j^{( J_{\mathrm{eff}} )}_{ {\mathrm{site B}},x}  $, and $ j^{( J_{\mathrm{eff}} )}_{ {\mathrm{site A}},z}  $ for $ -j^{( J_{\mathrm{eff}} )}_{ {\mathrm{site B}},z}  $. As a consequence, if a light field with $ F_{\mathrm{L}} $ and $ \theta_\text{ini}=\theta_0 $ produces $ j^{( J_{\mathrm{eff}} )}_{ {\mathrm{site A}},\perp} + j^{( J_{\mathrm{eff}} )}_{ {\mathrm{site B}},\perp} $ at time $ t $, it guarantees that a light field with $ F_{\mathrm{R}} $ of the same magnitude and $ \theta_\text{ini}=\theta_0 $ produces $ - j^{( J_{\mathrm{eff}} )}_{ {\mathrm{site A}},\perp} - j^{( J_{\mathrm{eff}} )}_{ {\mathrm{site B}},\perp} $ (i.e., of the same magnitude and opposite sign) at the same time $ t $, which is also numerically confirmed. 

\section{High-Frequency Expansion in Floquet Theory \label{sect:floquet}}
For interpreting photoinduced dynamics of pseudospins, a high-frequency expansion in quantum Floquet theory\cite{rahav_pra03,mananga_jcp11,goldman_prx14,eckardt_njp15,itin_prl15,bukov_ap15,mikami_prb16,yonemitsu_jpsj17b} turns out to be useful even if the frequencies used in the numerical calculations are not so high. Continuous waves are considered only in this section. For a circularly polarized light field, we use 
\begin{equation}
A_{a}(t) = \frac{c F_{\mathrm{L(R)}}}{\omega} \sin \omega t
\;, \label{eq:cw_circular_a}
\end{equation}
\begin{equation}
A_{b}(t) = \mp\frac{c F_{\mathrm{L(R)}}}{\omega} \cos \omega t
\;, \label{eq:cw_circular_b}
\end{equation}
\begin{equation}
J_m(ij) \equiv J_m \left( \frac{ea_{ij} F_{\mathrm{L(R)}}}{\hbar\omega}
\right) e^{\mp im\phi_{ij}}
\;, \label{eq:jm_circular}
\end{equation}
where $J_m(ij)$\cite{yonemitsu_jpsj17b} is used to define $H_m$ below, $ J_m(x) $ on the right-hand side is the $m$th-order Bessel function, $ a_{ij} = \mid \mbox{\boldmath $r$}_i-\mbox{\boldmath $r$}_j \mid $, and $ \phi_{ij} $ is the angle between $ \mbox{\boldmath $r$}_i-\mbox{\boldmath $r$}_j $ and a reference axis. Later, we often use $ H_m $ defined by 
\begin{equation}
H_m = -\sum_{ij} \vec{\mathbf{c}}_i^\dagger \left\{ \mathbf{T}_{ij} J_m(ij) \otimes \mathbb{I}_{2\times2} \right\} \vec{\mathbf{c}}_j
\;. \label{eq:mth_bessel}
\end{equation}
In the lowest order of the expansion (denoted by $ H_{\mathrm{F}}^{(1)} $ in Ref.~\citen{yonemitsu_jpsj17b}), the kinetic term of the Hamiltonian $ H_{\mathrm{hop}} $ [Eq.~(\ref{eq:kinetic})] is renormalized to be $ H_{m=0} $, while the other terms of the Hamiltonian are unaltered. 

In the second-lowest order, we have 
\begin{equation}
H_{\mathrm{F}}^{(2)} = \sum_{m \neq 0} \frac{H_{m}H_{-m}}{m \hbar \omega}
= \sum_{m>0} \frac{[ H_{m}, H_{-m} ]}{m \hbar \omega}
\;, \label{eq:expansion_2}
\end{equation}
which is written as 
\begin{equation}
H_{\mathrm{F}}^{(2)} =  \sum_{m>0,ijkabc\sigma} 
\frac{t^{ik}_{ac}t^{kj}_{cb}}{m \hbar \omega}
[J_{m}(ik) J_{-m}(kj) -J_{-m}(ik) J_{m}(kj)]
c_{i,a,\sigma}^\dagger c_{j,b,\sigma}
\;, \label{eq:expan2gen}
\end{equation}
where $ t^{ij}_{ab} $ denotes the transfer integral between orbital $a$ at site $i$ and orbital $b$ at site $j$. For a circularly polarized light field, Eq.~(\ref{eq:expan2gen}) is rewritten as 
\begin{equation}
H_{\mathrm{F}}^{(2)} = \sum_{m>0,ijkabc\sigma} 
\frac{t^{ik}_{ac}t^{kj}_{cb}}{m \hbar \omega}
J_{m}^2\left(\frac{ea F_{\mathrm{L(R)}}}{\hbar \omega}\right)(\mp2i)
\sin[m(\phi_{ik}-\phi_{jk})]
c_{i,a,\sigma}^\dagger c_{j,b,\sigma}
\;, \label{eq:expan2circ}
\end{equation}
where light fields of left-hand circular polarization and of right-hand circular polarization give the opposite signs. Note that, for a linearly polarized light field, Eq.~(\ref{eq:expan2gen}) becomes zero. 
For left-hand circular polarization, we proceed further. Since only next-nearest-neighbor sites $i$ and $j$ contribute to the summation in Eq.~(\ref{eq:expan2circ}), it is written as 
\begin{equation}
H_{\mathrm{F}}^{(2)} = 
H_{\mathrm{F}}^{(2,\mathrm{A})} + H_{\mathrm{F}}^{(2,\mathrm{B})}
\;, \label{eq:expan2AB}
\end{equation}
where sites $i$ and $j$ belong to sublattice A in $ H_{\mathrm{F}}^{(2,\mathrm{A})} $ and to sublattice B in $ H_{\mathrm{F}}^{(2,\mathrm{B})} $. With the Fourier transforms, $ H_{\mathrm{F}}^{(2,\mathrm{A})} $ is expressed as 
\begin{eqnarray}
H_{\mathrm{F}}^{(2,\mathrm{A})} & = & 
\sum_{m>0} \frac{1}{m \hbar \omega} \sum_{\mbox{\boldmath $k$}ab\sigma}
J_{m}^2\left(\frac{ea F_{\mathrm{L}}}{\hbar \omega}\right)(2i)
\sin\frac{2m\pi}{3} \nonumber \\
&\times& \left( 
e^{i\mbox{\boldmath $k$}\cdot\mbox{\boldmath $X_2$}} 
\mathbf{T}_1^{\mathrm{Y}} \mathbf{T}_1^{\mathrm{Z}} 
-e^{-i\mbox{\boldmath $k$}\cdot\mbox{\boldmath $X_2$}} 
\mathbf{T}_1^{\mathrm{Z}} \mathbf{T}_1^{\mathrm{Y}} 
\right. \nonumber \\ && 
+e^{i\mbox{\boldmath $k$}\cdot\mbox{\boldmath $Y_2$}}
\mathbf{T}_1^{\mathrm{Z}} \mathbf{T}_1^{\mathrm{X}} 
-e^{-i\mbox{\boldmath $k$}\cdot\mbox{\boldmath $Y_2$}} 
\mathbf{T}_1^{\mathrm{X}} \mathbf{T}_1^{\mathrm{Z}} 
\nonumber \\ && \left. 
+e^{i\mbox{\boldmath $k$}\cdot\mbox{\boldmath $Z_2$}}
\mathbf{T}_1^{\mathrm{X}} \mathbf{T}_1^{\mathrm{Y}} 
-e^{-i\mbox{\boldmath $k$}\cdot\mbox{\boldmath $Z_2$}} 
\mathbf{T}_1^{\mathrm{Y}} \mathbf{T}_1^{\mathrm{X}} 
\right)_{ab} 
c_{\mathrm{A},\mbox{\boldmath $k$},a,\sigma}^\dagger 
c_{\mathrm{A},\mbox{\boldmath $k$},b,\sigma}
\nonumber \\ & = & 
\sum_{m>0} \frac{1}{m \hbar \omega} \sum_{\mbox{\boldmath $k$}ab\sigma}
J_{m}^2\left(\frac{ea F_{\mathrm{L}}}{\hbar \omega}\right)(2i)
\sin\frac{2m\pi}{3} \nonumber \\
&\times& \left( 
\left[\mathbf{T}_1^{\mathrm{Y}}, \mathbf{T}_1^{\mathrm{Z}} \right]
\cos \mbox{\boldmath $k$}\cdot\mbox{\boldmath $X_2$} 
+i\left\{\mathbf{T}_1^{\mathrm{Y}}, \mathbf{T}_1^{\mathrm{Z}} \right\}
\sin \mbox{\boldmath $k$}\cdot\mbox{\boldmath $X_2$} 
\right. \nonumber \\ && 
+\left[ \mathbf{T}_1^{\mathrm{Z}}, \mathbf{T}_1^{\mathrm{X}} \right]
\cos \mbox{\boldmath $k$}\cdot\mbox{\boldmath $Y_2$}
+i\left\{ \mathbf{T}_1^{\mathrm{Z}}, \mathbf{T}_1^{\mathrm{X}} \right\}
\sin \mbox{\boldmath $k$}\cdot\mbox{\boldmath $Y_2$}
\nonumber \\ && \left. 
+\left[ \mathbf{T}_1^{\mathrm{X}}, \mathbf{T}_1^{\mathrm{Y}} \right]
\cos \mbox{\boldmath $k$}\cdot\mbox{\boldmath $Z_2$}
+i\left\{ \mathbf{T}_1^{\mathrm{X}}, \mathbf{T}_1^{\mathrm{Y}} \right\}
\sin \mbox{\boldmath $k$}\cdot\mbox{\boldmath $Z_2$}
\right)_{ab} 
c_{\mathrm{A},\mbox{\boldmath $k$},a,\sigma}^\dagger 
c_{\mathrm{A},\mbox{\boldmath $k$},b,\sigma}
\;, \label{eq:expan2A}
\end{eqnarray}
where the creation and annihilation operators act on sublattice A. 
The operator $ H_{\mathrm{F}}^{(2,\mathrm{B})} $ is given by reversing the momenta, i.e., $ \mbox{\boldmath $k$} \rightarrow -\mbox{\boldmath $k$} $, in the parentheses $\left( \cdots \right)_{ab}$ of Eq.~(\ref{eq:expan2A}) and 
$ c_{\mathrm{A},\mbox{\boldmath $k$},a,\sigma}^\dagger 
c_{\mathrm{A},\mbox{\boldmath $k$},b,\sigma} \rightarrow 
c_{\mathrm{B},\mbox{\boldmath $k$},a,\sigma}^\dagger 
c_{\mathrm{B},\mbox{\boldmath $k$},b,\sigma} $; 
the creation and annihilation operators now act on sublattice B. For right-hand circular polarization, the factor $(2i)$ in Eq.~(\ref{eq:expan2A}) is replaced by the factor $(-2i)$. In Eq.~(\ref{eq:expan2A}), the cosine terms are shown to be important below. This is in contrast to the one-orbital model on the honeycomb lattice (i.e., graphene) where the cosine terms are missing and the sine terms are responsible for the emergence of a Floquet topological insulator\cite{oka_prb09}. 

To make the implication of $ H_{\mathrm{F}}^{(2)} $ clear, we estimate it roughly. The term with $m$=1 is dominant when the light field is not so strong and the argument of the Bessel functions is small. Thus, the essential role of $ H_{\mathrm{F}}^{(2)} $ can be understood by approximating the $\mbox{\boldmath $k$}$-dependent terms in the parentheses by $\mbox{\boldmath $k$}$=0 and keeping the $m$=1 terms only: 
\begin{eqnarray}
H_{\mathrm{F}}^{(2)} & \simeq & \frac{1}{\hbar \omega} \sum_{iab\sigma}
J_{1}^2\left(\frac{ea F_{\mathrm{L}}}{\hbar \omega}\right)(2i)
\sin\frac{2\pi}{3} \nonumber \\
&\times& \left( 
 [\mathbf{T}_1^{\mathrm{Y}},\mathbf{T}_1^{\mathrm{Z}}]
+[\mathbf{T}_1^{\mathrm{Z}},\mathbf{T}_1^{\mathrm{X}}]
+[\mathbf{T}_1^{\mathrm{X}},\mathbf{T}_1^{\mathrm{Y}}] \right)_{ab}
c_{i,a,\sigma}^\dagger c_{i,b,\sigma}
\;. \label{eq:expan2approx}
\end{eqnarray}
Assuming the threefold symmetry, $ t_1 $=$ t^{\,\prime}_{1a} $=$ t^{\,\prime}_{1b} $, $ t_2 $=$ t^{\,\prime}_{2} $, $ t_3 $=$ t^{\,\prime}_{3} $, and $ t_4 $=$ t^{\,\prime}_{4a} $=$ t^{\,\prime}_{4b} $, we have 
\begin{eqnarray}
[\mathbf{T}_1^{\mathrm{Y}},\mathbf{T}_1^{\mathrm{Z}}] &=& 
-i\left[t_2(t_3-t_1)-t_4(t_2-t_4))\right](l_y+l_z) \nonumber \\ &&
-i\left[t_2^2-t_4^2-2t_4(t_3-t_1)\right]l_x
\;, \label{eq:commut_yz_approx}
\end{eqnarray}
\begin{eqnarray}
[\mathbf{T}_1^{\mathrm{Z}},\mathbf{T}_1^{\mathrm{X}}] &=&
-i\left[t_2(t_3-t_1)-t_4(t_2-t_4))\right](l_z+l_x) \nonumber \\ &&
-i\left[t_2^2-t_4^2-2t_4(t_3-t_1)\right]l_y
\;, \label{eq:commut_zx_approx}
\end{eqnarray}
\begin{eqnarray}
[\mathbf{T}_1^{\mathrm{X}},\mathbf{T}_1^{\mathrm{Y}}] &=&
-i\left[t_2(t_3-t_1)-t_4(t_2-t_4))\right](l_x+l_y) \nonumber \\ &&
-i\left[t_2^2-t_4^2-2t_4(t_3-t_1)\right]l_z
\;, \label{eq:commut_xy_approx}
\end{eqnarray}
with 
\begin{equation}
l_x = \begin{pmatrix}0&0&0\\0&0&-i\\0&i&0\\\end{pmatrix} \;, 
l_y = \begin{pmatrix}0&0&i\\0&0&0\\-i&0&0\\\end{pmatrix} \;, 
l_z = \begin{pmatrix}0&-i&0\\i&0&0\\0&0&0\\\end{pmatrix} 
\;; \label{eq:orb_xyz}
\end{equation}
thus, the commutators in Eq.~(\ref{eq:expan2approx}) are evaluated as 
\begin{eqnarray}
 [\mathbf{T}_1^{\mathrm{Y}},\mathbf{T}_1^{\mathrm{Z}}]
+[\mathbf{T}_1^{\mathrm{Z}},\mathbf{T}_1^{\mathrm{X}}]
+[\mathbf{T}_1^{\mathrm{X}},\mathbf{T}_1^{\mathrm{Y}}] &=& -i 
(t_2-t_4)[t_2-t_4+2(t_3-t_1)] \nonumber \\ & \times & 
\begin{pmatrix}
0 & -i & i \\ i & 0 & -i \\ -i & i & 0 \\
\end{pmatrix}
\;. \label{eq:commutator_approx}
\end{eqnarray}
Finally, we have 
\begin{eqnarray}
H_{\mathrm{F}}^{(2)} & \simeq & \frac{1}{\hbar \omega} 
J_{1}^2\left(\frac{ea F_{\mathrm{L}}}{\hbar \omega}\right)
\sqrt{3} (t_2-t_4)[t_2-t_4+2(t_3-t_1)] \nonumber \\
&\times& \sum_{i\sigma} 
\left( c_{i,yz,\sigma}^\dagger  \; 
c_{i,xz,\sigma}^\dagger  \; c_{i,xy,\sigma}^\dagger \right)
\begin{pmatrix}
0 & -i & i \\ i & 0 & -i \\ -i & i & 0 \\
\end{pmatrix}
\begin{pmatrix}
c_{i,yz,\sigma} \\ c_{i,xz,\sigma} \\ c_{i,xy,\sigma} \\
\end{pmatrix}
\;, \label{eq:expan2approx3fold_sym}
\end{eqnarray}
which implies the application of an effective magnetic field to the effective angular momenta. Because of the inequality $ (t_2-t_4)[t_2-t_4+2(t_3-t_1)] < 0$ owing to the opposite signs of the dominant transfer integrals $t_2$ and $t_3$, the effective magnetic field originating from $ H_{\mathrm{F}}^{(2)} $ points to the direction of $(1,1,1)$ (for left-hand circular polarization and to the direction of $(-1,-1,-1)$ for right-hand circular polarization). The factor appearing on the left of the summation symbol in Eq.~(\ref{eq:expan2approx3fold_sym}) becomes $-$2.6$\times$10$^{-4}$ ($-$2.4$\times$10$^{-2}$) for $F_{\mathrm{L}}$=0.02 ($F_{\mathrm{L}}$=0.2) and $\omega$=0.3 used later. These values in units of eV roughly correspond to 4.6 T (410 T). The emergence of an effective magnetic field without relying on spin-orbit coupling is theoretically proposed in a different context\cite{chono_prb20} in a similar manner to the present one. Note that a circularly-polarized-light-induced effective magnetic field in the direction perpendicular to the honeycomb lattice is derived in a different manner by considering ligand $p$ orbitals and ligand-mediated third-order hopping processes in a strong-coupling perturbation theory for the $J_{\mathrm{eff}}=\frac12$ pseudospins.\cite{sriram_arx,banerjee_prb22} On the other hand, the present field is derived by second-order hopping processes. 
Within the $J_{\mathrm{eff}}=\frac12$ subspace, Eq.~(\ref{eq:expan2approx3fold_sym}) is expressed as 
\begin{eqnarray}
H_{\mathrm{F}}^{(2,J_{\mathrm{eff}}=\frac12)} 
& \simeq & \frac{1}{\hbar \omega} 
J_{1}^2\left(\frac{ea F_{\mathrm{L}}}{\hbar \omega}\right)
\sqrt{3} (t_2-t_4)[t_2-t_4+2(t_3-t_1)] \nonumber \\
& \times & \frac{4}{3} \sum_{i} 
\begin{pmatrix}
p_{i,\uparrow}^\dagger & p_{i,\downarrow}^\dagger \\
\end{pmatrix}
\begin{pmatrix}
\frac{1}{2} & \frac{1-i}{2} \\ \frac{1+i}{2} & -\frac{1}{2} \\
\end{pmatrix}
\begin{pmatrix}
p_{i,\uparrow} \\ p_{i,\downarrow} \\
\end{pmatrix}
\;. \label{eq:expan2approx3fold_sym12}
\end{eqnarray}
Within the $J_{\mathrm{eff}}=\frac32$ subspace, Eq.~(\ref{eq:expan2approx3fold_sym}) leads to 
\begin{eqnarray}
H_{\mathrm{F}}^{(2,J_{\mathrm{eff}}=\frac32)} 
& \simeq & \frac{1}{\hbar \omega} 
J_{1}^2\left(\frac{ea F_{\mathrm{L}}}{\hbar \omega}\right)
\sqrt{3} (t_2-t_4)[t_2-t_4+2(t_3-t_1)] \nonumber \\
& \times & \frac{2}{3} \sum_{i} 
\begin{pmatrix}
q_{i,3/2}^\dagger & q_{i,1/2}^\dagger & q_{i,-1/2}^\dagger & q_{i,-3/2}^\dagger \\
\end{pmatrix} \nonumber \\ & \times & 
\begin{pmatrix}
\frac{3}{2} & \frac{\sqrt{3}}{2}(1-i) & 0 & 0 \\
\frac{\sqrt{3}}{2}(1+i) & \frac{1}{2} & 1-i & 0 \\
0 &  1+i & -\frac{1}{2} & \frac{\sqrt{3}}{2}(1-i) \\
0 & 0 & \frac{\sqrt{3}}{2}(1+i) & -\frac{3}{2} \\
\end{pmatrix}
\begin{pmatrix}
q_{i,3/2} \\ q_{i,1/2} \\ q_{i,-1/2} \\ q_{i,-3/2} \\
\end{pmatrix}
\;. \label{eq:expan2approx3fold_sym32}
\end{eqnarray}
The factor $\frac43$ in Eq.~(\ref{eq:expan2approx3fold_sym12}) and the factor $\frac23$ in Eq.~(\ref{eq:expan2approx3fold_sym32}) correspond to the Land\'e $g$-factors for $J_{\mathrm{eff}}=\frac12$ and $J_{\mathrm{eff}}=\frac32$, respectively, with $g_L=1$ and $g_S=0$ because the effective magnetic field is applied to the effective angular momenta of $L$=1. In the noninteracting case where we can calculate the photoinduced dynamics of pseudospin densities exactly, we confirm that the $(1,1,1)$ component of the moment is positive (negative) for left-hand (right-hand) circular polarization if spin-orbit coupling is dominant and $ (t_2-t_4)[t_2-t_4+2(t_3-t_1)] < 0$ is satisfied irrespective of $\omega < \frac32\lambda$ or $\omega > \frac32\lambda $ as long as the light field is not so strong. On the other hand, the sign is inverted if $ (t_2-t_4)[t_2-t_4+2(t_3-t_1)] > 0$ is satisfied. 

Because spin-orbit coupling [Eq.~(\ref{eq:spin_orbit})] operates within a site, its effect appears from the order of $\omega^{-2}$ (denoted by $ H_{\mathrm{F}}^{(3)} $ in Ref.~\citen{yonemitsu_jpsj17b}), which is given by 
\begin{equation}
H_{\mathrm{F},\mathrm{SO}}^{(3)} = \sum_{m \neq 0}
\frac{[ H_{-m},[ H_{\mathrm{SO}}, H_{m}]]}{2(m \hbar \omega)^2}
\;. \label{eq:spin_orbit3}
\end{equation}
Its meaning becomes clear when it is represented with $ p_{i,\sigma} $, $ p_{i,\sigma}^\dagger $ ($J_{\mathrm{eff}}=\frac12$ operators) and $ q_{i,J_z} $, $ q_{i,J_z}^\dagger $ ($J_{\mathrm{eff}}=\frac32$ operators). Here, we focus on the terms acting within the $J_{\mathrm{eff}}=\frac12$ subspace, $ H_{\mathrm{F},\mathrm{SO}}^{(3,J_{\mathrm{eff}}=\frac12)} $. The terms that act on pseudospins within the $J_{\mathrm{eff}}=\frac32$ subspace are shown in the Appendix. The operator $ H_{\mathrm{F},\mathrm{SO}}^{(3,J_{\mathrm{eff}}=\frac12)} $ turns out to be the sum of $ p_{i,\sigma}^\dagger p_{j,\tau} $ terms with $ i $=$ j $ or next-nearest-neighbor sites $ i $ and $ j $, where sites $ i $ and $ j $ belong to sublattice A, $ H_{\mathrm{F},\mathrm{SO}}^{(3,J_{\mathrm{eff}}=\frac12,\mathrm{A})} $, and  similar terms with sites $ i $ and $ j $ belonging to sublattice B, $ H_{\mathrm{F},\mathrm{SO}}^{(3,J_{\mathrm{eff}}=\frac12,\mathrm{B})} $. In momentum space, the operator $ H_{\mathrm{F},\mathrm{SO}}^{(3,J_{\mathrm{eff}}=\frac12,\mathrm{A})} $ is written as 
\begin{eqnarray}
H_{\mathrm{F},\mathrm{SO}}^{(3,J_{\mathrm{eff}}=\frac12,\mathrm{A})} & = & -
\sum_{\mbox{\boldmath $k$}} 
\left( p_{\mathrm{A},\mbox{\boldmath $k$},\uparrow}^\dagger 
p_{\mathrm{A},\mbox{\boldmath $k$},\downarrow}^\dagger \right) 
\nonumber \\ & \times &
\begin{pmatrix}
C(\mbox{\boldmath $k$})+\frac12 B^z_{\mathrm{eff}}(\mbox{\boldmath $k$}) &
\frac12 \left( B^x_{\mathrm{eff}}(\mbox{\boldmath $k$})
-i B^y_{\mathrm{eff}}(\mbox{\boldmath $k$}) \right) \\
\frac12 \left( B^x_{\mathrm{eff}}(\mbox{\boldmath $k$})
+i B^y_{\mathrm{eff}}(\mbox{\boldmath $k$}) \right) &
C(\mbox{\boldmath $k$})-\frac12 B^z_{\mathrm{eff}}(\mbox{\boldmath $k$}) \\
\end{pmatrix}
\begin{pmatrix}
p_{\mathrm{A},\mbox{\boldmath $k$},\uparrow} \\
p_{\mathrm{A},\mbox{\boldmath $k$},\downarrow} \\
\end{pmatrix}
\;, \label{eq:zeeman_12}
\end{eqnarray}
where the creation and annihilation operators act on sublattice A, and $ C(\mbox{\boldmath $k$}) $ is an even function of {\boldmath $k$}. The components of the effective magnetic field appearing above $ B^x_{\mathrm{eff}}(\mbox{\boldmath $k$}) $, $ B^y_{\mathrm{eff}}(\mbox{\boldmath $k$}) $, and $ B^z_{\mathrm{eff}}(\mbox{\boldmath $k$}) $ are inclusive of the Land\'e $g$-factor for $J_{\mathrm{eff}}=\frac12$ and odd functions of {\boldmath $k$}, which are given by 
\begin{equation}
B^{x,y,z}_{\mathrm{eff}}(\mbox{\boldmath $k$}) = 
\sum_{m \neq 0} \frac{-\lambda}{2(m\hbar\omega)^2} \times 
\left[ 
X_{m,\mathrm{eff}} (\mbox{\boldmath $k$}), 
Y_{m,\mathrm{eff}} (\mbox{\boldmath $k$}), 
Z_{m,\mathrm{eff}} (\mbox{\boldmath $k$}) \right]
\;, \label{eq:mag_field_12}
\end{equation}
with 
\begin{eqnarray}
X_{m,\mathrm{eff}} (\mbox{\boldmath $k$}) & = &
\left( -ie^{i\mbox{\boldmath $k$}\cdot\mbox{\boldmath $Y_2$}}
+ie^{-i\mbox{\boldmath $k$}\cdot\mbox{\boldmath $Y _2$}} \right) 
\left( J_{-m}(-\mbox{\boldmath $Z_1$}) J_{m}(\mbox{\boldmath $X_1$}) 
+ J_{m}(-\mbox{\boldmath $Z_1$}) J_{-m}(\mbox{\boldmath $X_1$}) 
\right) \nonumber \\
&& \times \left( 
-t_1 t^{\,\prime}_{2 } -t_2 t^{\,\prime}_{4b}
+t_3 t^{\,\prime}_{2 } +t_4 t^{\,\prime}_{1a}
-t_4 t^{\,\prime}_{1b} +t_4 t^{\,\prime}_{4a}
\right) \nonumber \\
&& + \left( -ie^{i\mbox{\boldmath $k$}\cdot\mbox{\boldmath $Z_2$}}
+ie^{-i\mbox{\boldmath $k$}\cdot\mbox{\boldmath $Z_2$}} \right) 
\left( J_{-m}(-\mbox{\boldmath $X_1$}) J_{m}(\mbox{\boldmath $Y_1$}) 
+ J_{m}(-\mbox{\boldmath $X_1$}) J_{-m}(\mbox{\boldmath $Y_1$}) 
\right) \nonumber \\
&& \times \left( 
- t^{\,\prime}_{1b} t^{\,\prime}_{2 } -t^{\,\prime}_{2 } t^{\,\prime}_{4a}
+ t^{\,\prime}_{2 } t^{\,\prime}_{3 } +t^{\,\prime}_{1b} t^{\,\prime}_{4b}
- t^{\,\prime}_{1a} t^{\,\prime}_{4b} +t^{\,\prime}_{4a} t^{\,\prime}_{4b}
\right) \nonumber \\
&& + \left( -ie^{i\mbox{\boldmath $k$}\cdot\mbox{\boldmath $X_2$}}
+ie^{-i\mbox{\boldmath $k$}\cdot\mbox{\boldmath $X_2$}} \right) 
\left( J_{-m}(-\mbox{\boldmath $Y_1$}) J_{m}(\mbox{\boldmath $Z_1$}) 
+ J_{m}(-\mbox{\boldmath $Y_1$}) J_{-m}(\mbox{\boldmath $Z_1$}) 
\right) \nonumber \\
&& \times \left( 
 t_1 t^{\,\prime}_{4b} +t_4 t^{\,\prime}_{1b}
-t_3 t^{\,\prime}_{4b} -t_4 t^{\,\prime}_{3 }
+t_2 t^{\,\prime}_{2 } -t_4 t^{\,\prime}_{4a}
\right) 
\;, \label{eq:mag_f_x_eff}
\end{eqnarray}
\begin{eqnarray}
Y_{m,\mathrm{eff}} (\mbox{\boldmath $k$}) & = &
\left( -ie^{i\mbox{\boldmath $k$}\cdot\mbox{\boldmath $Z_2$}}
+ie^{-i\mbox{\boldmath $k$}\cdot\mbox{\boldmath $Z_2$}} \right) 
\left( J_{-m}(-\mbox{\boldmath $X_1$}) J_{m}(\mbox{\boldmath $Y_1$}) 
+ J_{m}(-\mbox{\boldmath $X_1$}) J_{-m}(\mbox{\boldmath $Y_1$}) 
\right) \nonumber \\
&& \times \left( 
- t^{\,\prime}_{1b} t^{\,\prime}_{2 } -t^{\,\prime}_{2 } t^{\,\prime}_{4a}
+ t^{\,\prime}_{2 } t^{\,\prime}_{3 } +t^{\,\prime}_{1b} t^{\,\prime}_{4b}
- t^{\,\prime}_{1a} t^{\,\prime}_{4b} +t^{\,\prime}_{4a} t^{\,\prime}_{4b}
\right) \nonumber \\
&& + \left( -ie^{i\mbox{\boldmath $k$}\cdot\mbox{\boldmath $X_2$}}
+ie^{-i\mbox{\boldmath $k$}\cdot\mbox{\boldmath $X_2$}} \right) 
\left( J_{-m}(-\mbox{\boldmath $Y_1$}) J_{m}(\mbox{\boldmath $Z_1$}) 
+ J_{m}(-\mbox{\boldmath $Y_1$}) J_{-m}(\mbox{\boldmath $Z_1$}) 
\right) \nonumber \\
&& \times \left( 
-t_1 t^{\,\prime}_{2 } -t_2 t^{\,\prime}_{4b}
+t_3 t^{\,\prime}_{2 } +t_4 t^{\,\prime}_{1a}
-t_4 t^{\,\prime}_{1b} +t_4 t^{\,\prime}_{4a}
\right) \nonumber \\
&& + \left( -ie^{i\mbox{\boldmath $k$}\cdot\mbox{\boldmath $Y_2$}}
+ie^{-i\mbox{\boldmath $k$}\cdot\mbox{\boldmath $Y _2$}} \right) 
\left( J_{-m}(-\mbox{\boldmath $Z_1$}) J_{m}(\mbox{\boldmath $X_1$}) 
+ J_{m}(-\mbox{\boldmath $Z_1$}) J_{-m}(\mbox{\boldmath $X_1$}) 
\right) \nonumber \\
&& \times \left( 
 t_1 t^{\,\prime}_{4b} +t_4 t^{\,\prime}_{1b}
-t_3 t^{\,\prime}_{4b} -t_4 t^{\,\prime}_{3 }
+t_2 t^{\,\prime}_{2 } -t_4 t^{\,\prime}_{4a}
\right) 
\;, \label{eq:mag_f_y_eff}
\end{eqnarray}
\begin{eqnarray}
Z_{m,\mathrm{eff}} (\mbox{\boldmath $k$}) & = & 
\left[ \left( -ie^{i\mbox{\boldmath $k$}\cdot\mbox{\boldmath $X_2$}}
+ie^{-i\mbox{\boldmath $k$}\cdot\mbox{\boldmath $X_2$}} \right) 
\left( J_{-m}(-\mbox{\boldmath $Y_1$}) J_{m}(\mbox{\boldmath $Z_1$}) 
+ J_{m}(-\mbox{\boldmath $Y_1$}) J_{-m}(\mbox{\boldmath $Z_1$}) \right) 
\right. \nonumber \\
&& \left. +\left( -ie^{i\mbox{\boldmath $k$}\cdot\mbox{\boldmath $Y_2$}}
+ie^{-i\mbox{\boldmath $k$}\cdot\mbox{\boldmath $Y_2$}} \right) 
\left( J_{-m}(-\mbox{\boldmath $Z_1$}) J_{m}(\mbox{\boldmath $X_1$}) 
+ J_{m}(-\mbox{\boldmath $Z_1$}) J_{-m}(\mbox{\boldmath $X_1$}) \right) 
\right] \nonumber \\
&& \times \left( 
-t_2 t^{\,\prime}_{1a} -t_4 t^{\,\prime}_{2 } 
+t_2 t^{\,\prime}_{3 } +t_4 t^{\,\prime}_{4b} 
\right) \nonumber \\
&& +\left( -ie^{i\mbox{\boldmath $k$}\cdot\mbox{\boldmath $Z_2$}}
+ie^{-i\mbox{\boldmath $k$}\cdot\mbox{\boldmath $Z_2$}} \right) 
\left( J_{-m}(-\mbox{\boldmath $X_1$}) J_{m}(\mbox{\boldmath $Y_1$}) 
+ J_{m}(-\mbox{\boldmath $X_1$}) J_{-m}(\mbox{\boldmath $Y_1$}) 
\right) \nonumber \\
&& \times \left( 
2t^{\,\prime}_{1a} t^{\,\prime}_{4a} -2t^{\,\prime}_{3 } t^{\,\prime}_{4a} 
+t^{\,\prime}_{2 } t^{\,\prime}_{2 } - t^{\,\prime}_{4b} t^{\,\prime}_{4b} 
\right)
\;. \label{eq:mag_f_z_eff}
\end{eqnarray}
Here, the arguments of $J_{\pm m}$ [defined in Eq.~(\ref{eq:jm_circular})] imply that $ \mbox{\boldmath $r$}_i-\mbox{\boldmath $r$}_j $ is one of $ \pm \mbox{\boldmath $X_1$} $, $ \pm \mbox{\boldmath $Y_1$} $, and $ \pm \mbox{\boldmath $Z_1$} $ indicated in Fig.~\ref{fig:honeycomb}. The vectors {\boldmath $X_2$}, {\boldmath $Y_2$}, and {\boldmath $Z_2$} in the exponents are shown in Fig.~\ref{fig:honeycomb}. 
The operator $ H_{\mathrm{F},\mathrm{SO}}^{(3,J_{\mathrm{eff}}=\frac12,\mathrm{B})} $ is given by reversing the momenta, i.e., $ \mbox{\boldmath $k$} \rightarrow -\mbox{\boldmath $k$} $, in the 2$\times$2 matrix of Eq.~(\ref{eq:zeeman_12}) and $ p_{\mathrm{A},\mbox{\boldmath $k$},\sigma} \rightarrow p_{\mathrm{B},\mbox{\boldmath $k$},\sigma} $, $ p_{\mathrm{A},\mbox{\boldmath $k$},\sigma}^\dagger \rightarrow p_{\mathrm{B},\mbox{\boldmath $k$},\sigma}^\dagger $ for $ \sigma=\uparrow,\downarrow $; the creation and annihilation operators now act on sublattice B. 

The effective magnetic fields given by Eq.~(\ref{eq:mag_field_12}) have the properties originating from the symmetry mentioned in Sect.~\ref{sect:reflection}: the simultaneous exchanges of vectors {\boldmath $X_1$} for {\boldmath $Y_1$}, {\boldmath $X_2$} for $ -\mbox{\boldmath $Y_2$} $, and {\boldmath $Z_2$} for $ -\mbox{\boldmath $Z_2$} $ lead to the exchanges of $ B^x_{\mathrm{eff}}(\mbox{\boldmath $k$}) $ for $ -B^y_{\mathrm{eff}}(\mbox{\boldmath $k$}) $ and $ B^z_{\mathrm{eff}}(\mbox{\boldmath $k$}) $ for $ -B^z_{\mathrm{eff}}(\mbox{\boldmath $k$}) $. In the threefold-symmetric case of $ t_1 $=$ t^{\,\prime}_{1a} $=$ t^{\,\prime}_{1b} $, $ t_2 $=$ t^{\,\prime}_{2} $, $ t_3 $=$ t^{\,\prime}_{3} $, and $ t_4 $=$ t^{\,\prime}_{4a} $=$ t^{\,\prime}_{4b} $, the counterclockwise rotation through 120 degrees ({\boldmath $X_i$}$\rightarrow${\boldmath $Y_i$}$\rightarrow${\boldmath $Z_i$}$\rightarrow${\boldmath $X_i$} with $i$=1 and 2) leads to $ B^x_{\mathrm{eff}}(\mbox{\boldmath $k$}) \rightarrow B^y_{\mathrm{eff}}(\mbox{\boldmath $k$}) \rightarrow B^z_{\mathrm{eff}}(\mbox{\boldmath $k$}) \rightarrow B^x_{\mathrm{eff}}(\mbox{\boldmath $k$}) $. If $t_1$=$t_3$ and $t_2$=$t_4$ are additionally satisfied, the effective magnetic fields vanish. According to Refs.~\citen{rau_prl14} and \citen{winter_prb16}, the nearest-neighbor Kitaev coupling also vanishes in this case. 

The present system is almost, but not quite, threefold-symmetric. 
When the values of the transfer integrals (that satisfy $ t_1 \simeq t^{\,\prime}_{1a} \simeq t^{\,\prime}_{1b} $, $ t_2 \simeq t^{\,\prime}_{2} $, $ t_3 \simeq t^{\,\prime}_{3} $, and $ t_4 \simeq t^{\,\prime}_{4a} \simeq t^{\,\prime}_{4b} $) are substituted into the equations above, the factor $ \left( 
2t^{\,\prime}_{1a} t^{\,\prime}_{4a} -2t^{\,\prime}_{3 } t^{\,\prime}_{4a} 
+t^{\,\prime}_{2 } t^{\,\prime}_{2 } - t^{\,\prime}_{4b} t^{\,\prime}_{4b} 
\right) $ in Eq.~(\ref{eq:mag_f_z_eff}) and similar factors in Eqs.~(\ref{eq:mag_f_x_eff}) and (\ref{eq:mag_f_y_eff}) are positive, while the factor $ \left( 
-t_2 t^{\,\prime}_{1a} -t_4 t^{\,\prime}_{2 } 
+t_2 t^{\,\prime}_{3 } +t_4 t^{\,\prime}_{4b} 
\right) $ in Eq.~(\ref{eq:mag_f_z_eff}) and similar factors in Eqs.~(\ref{eq:mag_f_x_eff}) and (\ref{eq:mag_f_y_eff}) are negative. The relative signs of these factors turn out to be important. If the hole density in momentum space becomes disproportionate along the $ \mbox{\boldmath $k$}\cdot\mbox{\boldmath $X_2$} $-axis, the effect of $ B^x_{\mathrm{eff}}(\mbox{\boldmath $k$}) $ becomes large. If it becomes disproportionate along the $ \mbox{\boldmath $k$}\cdot\mbox{\boldmath $Y_2$} $-axis, the effect of $ B^y_{\mathrm{eff}}(\mbox{\boldmath $k$}) $ becomes large. If it becomes disproportionate along the $ \mbox{\boldmath $k$}\cdot\mbox{\boldmath $Z_2$} $-axis, the effect of $ B^z_{\mathrm{eff}}(\mbox{\boldmath $k$}) $ becomes large. If the direction of the current density is rotated counterclockwise as time advances during the photoexcitation of left-hand circular polarization, the effective magnetic field at site A originating from $ H_{\mathrm{F},\mathrm{SO}}^{(3)} $ is rotated roughly as $x$ $\rightarrow$ $y$ $\rightarrow$ $z$ directions nearly normal to the (1,1,1) direction. During the same time period, the effective magnetic field at site B originating from $ H_{\mathrm{F},\mathrm{SO}}^{(3)} $ is rotated roughly as $-x$ $\rightarrow$ $-y$ $\rightarrow$ $-z$ directions. These fields at sites A and B are almost antiparallel as long as the distribution of the hole density at site A and that at site B are similar in momentum space. This character of the effective magnetic fields is numerically confirmed when the light field is weak, as shown later. Note that $ H_{\mathrm{F},\mathrm{SO}}^{(3)} $ is responsible for the stroboscopic time evolution, and the relationship between $ \mbox{\boldmath $B$}_{\mathrm{eff}}(\mbox{\boldmath $k$}) $ and the momentum distribution of holes  should be understood to be a time-averaged one. Because the momentum distribution of holes varies according to the polarization of the light field, the effect of $ \mbox{\boldmath $B$}_{\mathrm{eff}}(\mbox{\boldmath $k$}) $ should also vary on the same timescale. 

\section{Optical Conductivity}
\begin{figure}
\includegraphics[height=13.6cm]{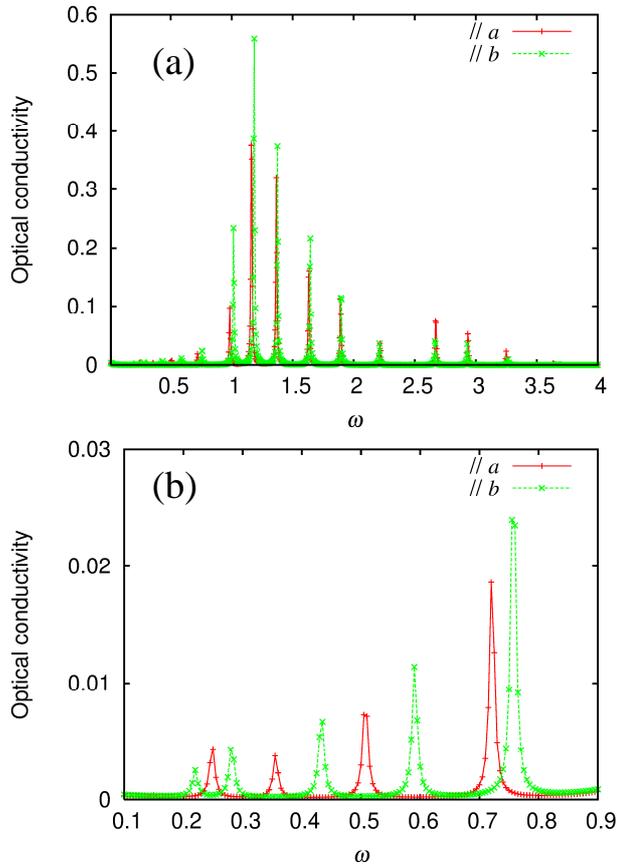}
\caption{(Color online) 
Optical conductivity spectra with polarization parallel to $a$- and $b$-axes in the ground state. In (b), the spectra below the Mott gap in (a) are enlarged. 
\label{fig:opt_cond}}
\end{figure}
It is now recognized that $\alpha$-RuCl$_3$ is a Mott insulator with a fundamental optical gap (the so-called Mott gap) of 1.1 eV. Thus, the observed peak at 1.2 eV is interpreted as an intersite $dd$ transition involving $t_{2g}$ orbitals of neighboring Ru ions. The peak at 2.0 eV is also interpreted as an intersite $dd$ transition in Refs.~\citen{sandilands_prb16a} and \citen{sandilands_prb16b}, while it is interpreted as an excitation into an $e_g$ orbital in Ref.~\citen{koitzsch_prl16}. At energies higher than about 3 eV, there are contributions from Ru $e_g$ orbitals or Cl $p$ orbitals, which are not taken into account in the present calculation. In the calculated optical conductivity spectra shown in Fig.~\ref{fig:opt_cond}(a), where energies are expressed in units of eV, qualitative characteristics of the experimentally observed spectra are reproduced. A main peak is located at around 1.2 eV. The Mott gap appears to be about 1 eV. 

Below the Mott gap, three infrared peaks are observed at about 0.3 eV, 0.5 eV, and 0.7 eV.\cite{sandilands_prb16a,warzanowski_prr20} In the latest report,\cite{warzanowski_prr20} they are interpreted as phonon-assisted excitations of single and double spin-orbit excitons and the direct excitation of a triple spin-orbit exciton. Multiple spin-orbit excitons are examined in detail by combining Raman spectroscopy with exact diagonalization calculations for a six-site system in Ref.~\citen{lee_npjqm21}. Some infrared peaks below the Mott gap are found in the calculated spectra shown in Fig.~\ref{fig:opt_cond}(b). If we use small parameter values for $ H_{\mathrm{hop}} $ and set $ H_{\mathrm{CF}} $=0 even in our calculation for the model without phonons, we find peaks at $\frac32 \lambda$, which corresponds to a transition between $J_{\mathrm{eff}}=\frac12$ and $J_{\mathrm{eff}}=\frac32$ states, $2\times\frac32 \lambda$, and $3\times\frac32 \lambda$, although their oscillation strengths are quite small. They would correspond to single, double, and triple spin-orbit excitons. Because the present system has two sites in a unit cell, both optically allowed and optically forbidden transitions could be constructed without the help of phonons, although the oscillator strengths of some transitions would be enhanced by phonons. 
In fact, in the experiment, the spectral weights of the three infrared peaks are not small even at the lowest temperature;\cite{warzanowski_prr20} thus, they would survive even if phonons were absent. In reality, however, the parameter values for $ H_{\mathrm{hop}} $ are comparable to $\lambda$. The number of holes in the $J_{\mathrm{eff}}=\frac32$ subspace is not a good quantum number. This fact makes the situation for the peak energies, the oscillator strengths, and the number of peaks itself quite different from those in the large-$ \lambda $ case. This would be the reason why the assignment of the infrared peaks has been controversial. It is difficult, by our calculation based on the small cluster that omits hopping to the second-nearest and third-nearest neighbors, to quantitatively reproduce the infrared peaks. It is numerically true that the spectral shape below the Mott gap is sensitive to the value of $\lambda$. The present value $ \lambda $=0.15 eV taken from Ref.~\citen{winter_prb16} is close to the value $ \lambda $=0.16 eV evaluated in Ref.~\citen{warzanowski_prr20}. 

\section{Photoinduced Pseudospin Dynamics}
We discuss the dynamics of pseudospin densities during the application of a circularly polarized light field. The frequency of the light field is set below the Mott gap. Although the expansion developed in Sect.~\ref{sect:floquet} is justified when the frequency is high, we expect that the properties originating from the symmetry aspects are qualitatively unaltered even when the frequency is set to be low and the electronic transition processes contributing to the effective magnetic fields are weighted differently from the high-frequency case. Then, in interpreting the photoinduced dynamics, we will refer to the high-frequency expansion and see how they are similar to or different from the behavior expected by this expansion. Although the system treated here is purely two-dimensional, the model parameters used here are derived for the three-dimensional system.\cite{winter_prb16} Since the model parameters are not expected to be qualitatively changed even for monolayer to few-layer systems,\cite{lee_npjqm21} the present results would be qualitatively valid even for such systems. Hereafter, energies are given in units of eV. 

\begin{figure}
\includegraphics[height=23.6cm]{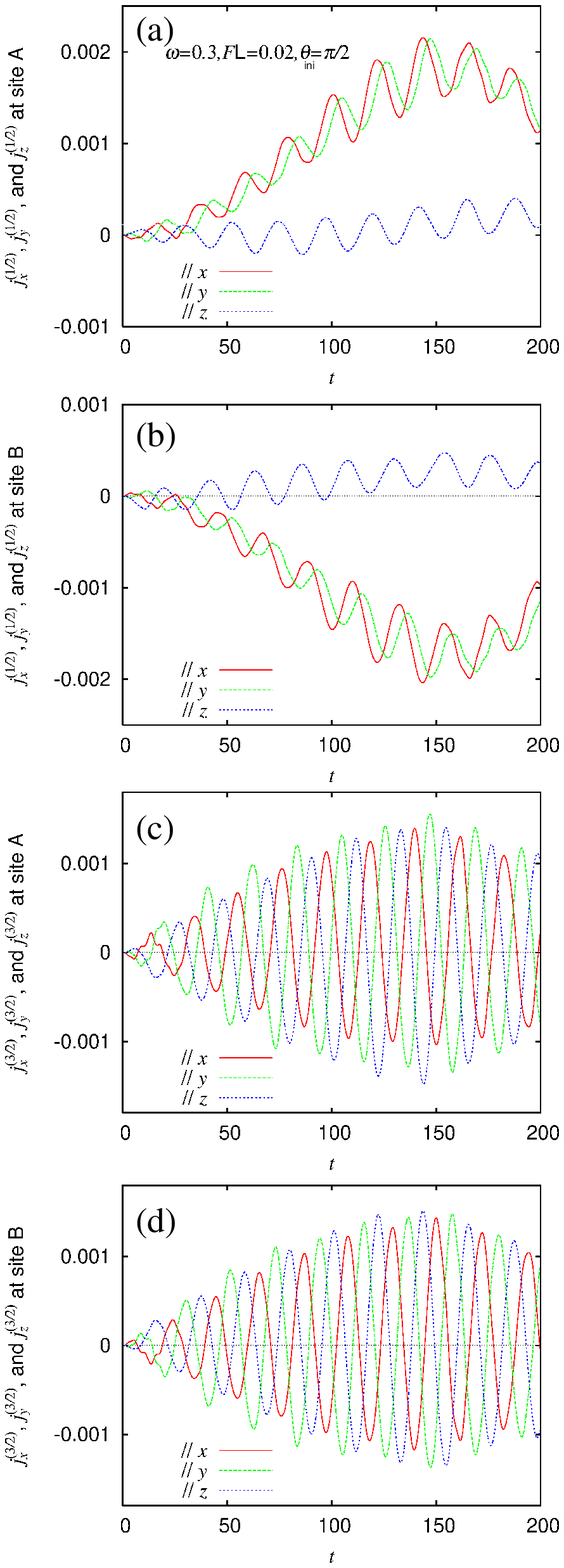}
\caption{(Color online) 
Time profiles of $x$, $y$, and $z$ components of (a, b) $J_{\mathrm{eff}}=\frac12$ pseudospins and (c, d) $J_{\mathrm{eff}}=\frac32$ pseudospins (a, c) at site A and (b, d) at site B, when excited by a circularly polarized light field with $\omega$=0.3, $F_{\mathrm{L}}$=0.02, and $\theta_\text{ini}$=$\pi/2$. The duration of photoexcitation is $0<t<t_{\mathrm{off}}\simeq$147 ($n$=7). 
\label{fig:weak_f_ps_AB}}
\end{figure}
As described in Sect.~\ref{sect:floquet}, when the direction of the current density is rotated counterclockwise during the application of a light field with left-hand circular polarization, the effective magnetic field originating from $ H_{\mathrm{F},\mathrm{SO}}^{(3)} $ at site A and that at site B are antiparallel and they are rotated counterclockwise within the $a$-$b$ plane. The pseudospins would behave in a similar manner. 
When the light field is weak, this behavior is indeed realized, as shown in Fig.~\ref{fig:weak_f_ps_AB} for $\omega$=0.3, which roughly corresponds to a single spin-orbit exciton. Later the $\omega$ dependence of the dynamics is discussed, but this behavior is always realized and independent of $\omega$ as long as the light field is weak. 
The time profiles of the $x$, $y$, and $z$ components of the $J_{\mathrm{eff}}=\frac12$ pseudospin densities at site A ($j^{(1/2)}_{\mathrm{site A},x}$, $j^{(1/2)}_{\mathrm{site A},y}$, and $j^{(1/2)}_{\mathrm{site A},z}$) and at site B ($j^{(1/2)}_{\mathrm{site B},x}$, $j^{(1/2)}_{\mathrm{site B},y}$, and $j^{(1/2)}_{\mathrm{site B},z}$) are shown for $ \theta_\text{ini}=\frac{\pi}{2} $ in Figs.~\ref{fig:weak_f_ps_AB}(a) and \ref{fig:weak_f_ps_AB}(b), respectively. Those of the $J_{\mathrm{eff}}=\frac32$ pseudospin densities at site A ($j^{(3/2)}_{\mathrm{site A},x}$, $j^{(3/2)}_{\mathrm{site A},y}$, and $j^{(3/2)}_{\mathrm{site A},z}$) and at site B ($j^{(3/2)}_{\mathrm{site B},x}$, $j^{(3/2)}_{\mathrm{site B},y}$, and $j^{(3/2)}_{\mathrm{site B},z}$) are shown under the same conditions in Figs.~\ref{fig:weak_f_ps_AB}(c) and \ref{fig:weak_f_ps_AB}(d), respectively. 
The $x$, $y$, and $z$ components show local maxima in this order as time advances, and their short-time averages (over the period of $ 2\pi/\omega $) are very small, especially for $J_{\mathrm{eff}}=\frac32$. This indicates that the $J_{\mathrm{eff}}=\frac32$ pseudospins at sites A and B are perpendicular to the (1,1,1) axis and rotate counterclockwise. This is what we expect from $ H_{\mathrm{F},\mathrm{SO}}^{(3)} $ in quantum Floquet theory. The behavior of the $J_{\mathrm{eff}}=\frac12$ pseudospins is roughly similar to that of the $J_{\mathrm{eff}}=\frac32$ pseudospins, but there is some difference presumably because the $J_{\mathrm{eff}}=\frac12$ pseudospins are directly influenced by the abrupt application of the light field at $t$=0. More precisely, the short-time averages of these components deviate from zero, and the motion of each pseudospin is not pure rotation around the (1,1,1) axis. For $ \theta_\text{ini}=\frac{\pi}{2} $ ($ \mbox{\boldmath $E$} \parallel \mbox{\boldmath $b$} $ at $t$=0) in fact, the short-time average of the $J_{\mathrm{eff}}=\frac12$ pseudospin at site A deviates roughly to the (1,1,0) direction, and that at site B deviates roughly to the ($-1$,$-1$,0) direction. For $ \theta_\text{ini}=0 $ ($ \mbox{\boldmath $E$} \parallel \mbox{\boldmath $a$} $ at $t$=0) on the other hand, the short-time average of the $J_{\mathrm{eff}}=\frac12$ pseudospin at site A deviates roughly to the ($-1$,1,0) direction, and that at site B deviates roughly to the (1,$-1$,0) direction (not shown). In any case, the pseudospins at sites A and B are almost antiparallel for both $J_{\mathrm{eff}}=\frac12$ and $J_{\mathrm{eff}}=\frac32$. As a consequence, the pseudospin densities averaged over sites A and B hardly grow as long as the light field is weak. Because the behavior of the $J_{\mathrm{eff}}=\frac32$ pseudospins is similar to that of the $J_{\mathrm{eff}}=\frac12$ pseudospins, we hereafter show only the $J_{\mathrm{eff}}=\frac12$ pseudospins, which are directly related to the magnetic moments. Thus, the pseudospins denote the $J_{\mathrm{eff}}=\frac12$ pseudospins unless stated otherwise. 

\begin{figure}
\includegraphics[height=13.6cm]{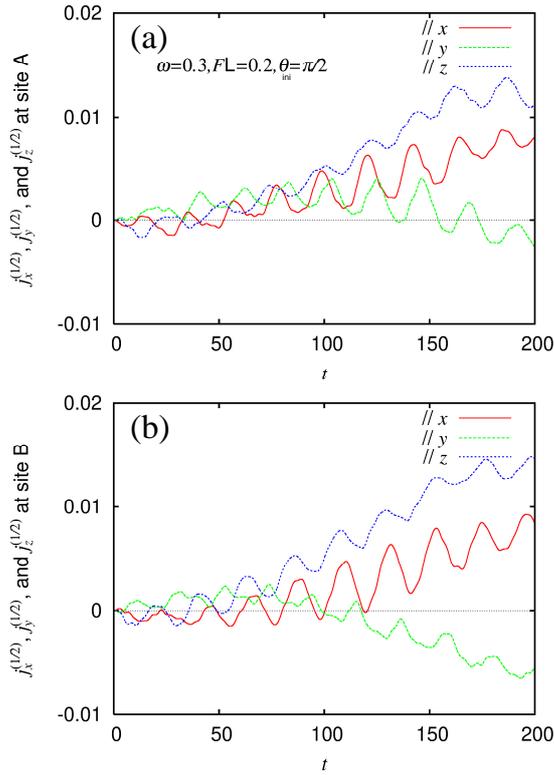}
\caption{(Color online) 
Time profiles of $x$, $y$, and $z$ components of $J_{\mathrm{eff}}=\frac12$ pseudospins (a) at site A and (b) at site B, when excited by a circularly polarized light field with $\omega$=0.3, $F_{\mathrm{L}}$=0.2, and $\theta_\text{ini}$=$\pi/2$. The duration of photoexcitation is $0<t<t_{\mathrm{off}}\simeq$147 ($n$=7). 
\label{fig:strong_f_ps_AB}}
\end{figure}
When the light field is increased, the antiparallel property of the pseudospins at sites A and B breaks down, as shown in Fig.~\ref{fig:strong_f_ps_AB}. The time profiles of the $x$, $y$, and $z$ components of the pseudospin densities at site A ($j^{(1/2)}_{\mathrm{site A},x}$, $j^{(1/2)}_{\mathrm{site A},y}$, and $j^{(1/2)}_{\mathrm{site A},z}$) and at site B ($j^{(1/2)}_{\mathrm{site B},x}$, $j^{(1/2)}_{\mathrm{site B},y}$, and $j^{(1/2)}_{\mathrm{site B},z}$) are shown for $\theta_\text{ini}$=$\pi/2$ in Figs.~\ref{fig:strong_f_ps_AB}(a) and \ref{fig:strong_f_ps_AB}(b), respectively. At both sites, the $x$, $y$, and $z$ components show local maxima in this order as time advances, and their short-time averages mainly increase with time until the light field is switched off. Therefore, the $\perp$ component of the averaged pseudospin density grows and attains a positive value. This finding is consistent with the effective magnetic field originating from $ H_{\mathrm{F}}^{(2)} $, which is discussed in Sect.~\ref{sect:floquet}. In fact, when the signs of $t_3$ and $ t^{\,\prime}_{3} $ are reversed to have $ (t_2-t_4)[t_2-t_4+2(t_3-t_1)] > 0$, the effective magnetic field originating from $ H_{\mathrm{F}}^{(2)} $ is reversed, and the sign of the $\perp$ component is numerically confirmed to be reversed. 

\begin{figure}
\includegraphics[height=20.4cm]{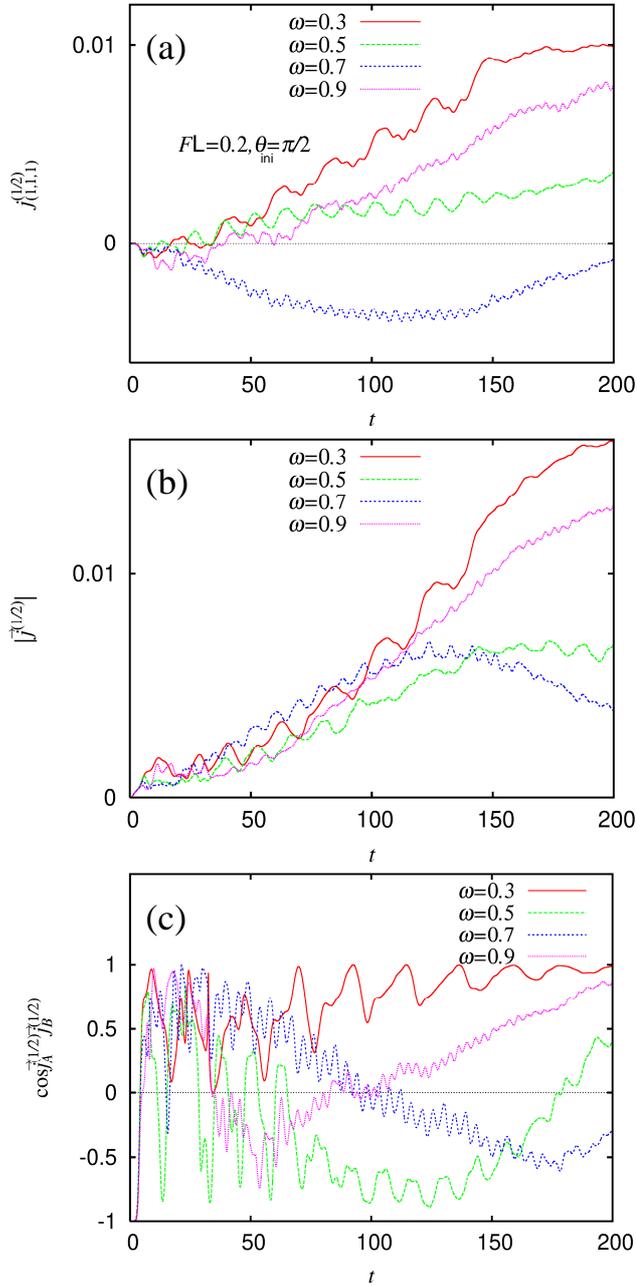}
\caption{(Color online) 
Time profiles of (a) averaged $\perp$ component and (b) averaged magnitude of $J_{\mathrm{eff}}=\frac12$ pseudospins, and that of (c) cosine of angle between $J_{\mathrm{eff}}=\frac12$ pseudospins at sites A and B, when excited by a circularly polarized light field with $F_{\mathrm{L}}$=0.2 and $\theta_\text{ini}$=$\pi/2$. The duration of photoexcitation is $0<t<t_{\mathrm{off}}\simeq$147 ($n$=7) for $\omega$=0.3, 138 ($n$=11) for $\omega$=0.5, 144 ($n$=16) for $\omega$=0.7, and 140 ($n$=20) for $\omega$=0.9. 
\label{fig:strong_f_ps_w_dep}}
\end{figure}
For the same light field at $t$=0 as that in Fig.~\ref{fig:strong_f_ps_AB} ($F_{\mathrm{L}}$=0.2 and $\theta_\text{ini}$=$\pi/2$), the $\omega$ dependence of the pseudospin dynamics is shown in Fig.~\ref{fig:strong_f_ps_w_dep}. The duration of photoexcitation $ t_{\mathrm{off}} $($\simeq$140) is set to be a multiple of the period $ 2\pi/\omega $, which depends on $\omega$. The averaged $\perp$ component of the pseudospin densities $\frac12 (j^{(1/2)}_{\mathrm{site A},\perp}+j^{(1/2)}_{\mathrm{site B},\perp})$ evolves with time, as shown in Fig.~\ref{fig:strong_f_ps_w_dep}(a). It obtains a positive value when $\omega$ is small (at least for $\omega \leq$0.6; not shown), and its growth basically stops at $ t = t_{\mathrm{off}} $ because the effective magnetic fields disappear after photoexcitation. If we could treat larger systems, some degrees of freedom other than the pseudospin ones would act as a heat bath, and the $\perp$ component would decay. The fine structures on the timescale below 6 (corresponding to the energy scale above 1 eV) shown hereafter would originate from the discreteness of the excitation spectra [Fig.~\ref{fig:opt_cond}(a)]; thus, they would disappear in the thermodynamic limit. The presented $\omega$-dependent behavior during the photoexcitation would be related to the characteristics of in-gap states below the Mott gap, i.e., it is presumably because different types of spin-orbit excitons are produced. This is shortly explained. Among the values of $\omega$ used in this calculation, the largest $\perp$ component is realized with $\omega$=0.3, which roughly corresponds to a single spin-orbit exciton; thus, we mainly focus on the dynamics in this case. 

\begin{figure}
\includegraphics[height=13.6cm]{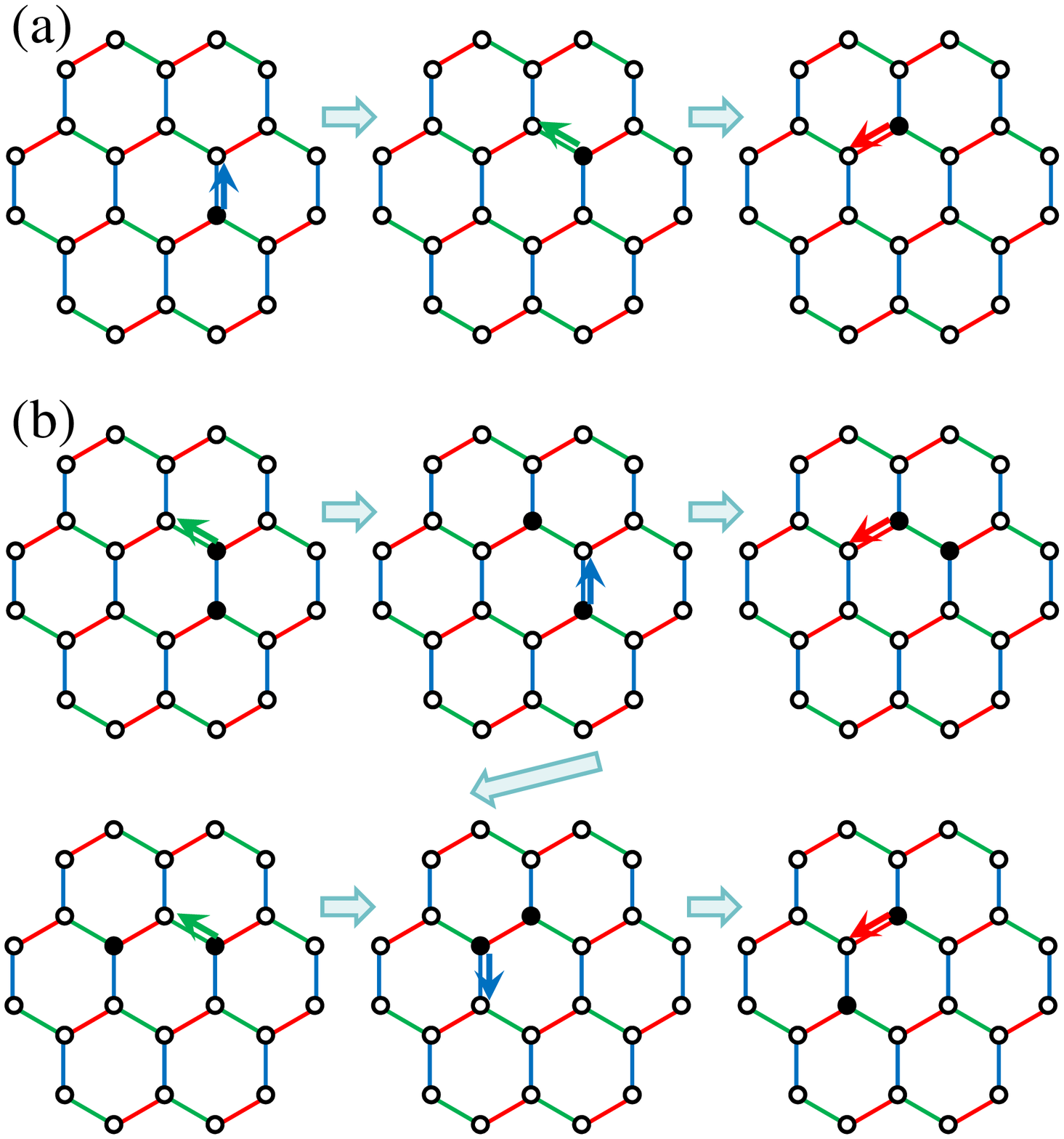}
\caption{(Color online) 
Schematic hopping processes for (a) single and (b) double spin-orbit excitons. The black circles show the positions of holes in the $J_{\mathrm{eff}}=\frac32$ subspace. 
\label{fig:honeycomb_hoppings}}
\end{figure}
For $\omega$=0.7, a double spin-orbit exciton would be produced. We expect its motion to be quite different from that of a single spin-orbit exciton during the photoexcitation of left-hand circular polarization, as shown in Fig.~\ref{fig:honeycomb_hoppings}. In both cases, the overall direction of the current density is rotated counterclockwise. For a single spin-orbit exciton, the hole in the $J_{\mathrm{eff}}=\frac32$ subspace is simply pushed in the direction of the electric field at the time, as those in the $J_{\mathrm{eff}}=\frac12$ subspace. On average, it hops on the Z$_1$ bond, on the Y$_1$ bond, and then on the X$_1$ bond as time advances [Fig.~\ref{fig:honeycomb_hoppings}(a)]. For a double spin-orbit exciton, however, the holes in the $J_{\mathrm{eff}}=\frac32$ subspace cannot occupy a single site; thus, the hopping processes are necessarily correlated. If the holes in the $J_{\mathrm{eff}}=\frac32$ subspace are initially located along a Z$_1$ bond, a hole hops on the Y$_1$ bond and then the other hops on the Z$_1$ bond to form a double spin-orbit exciton on the Y$_1$ bond. The next motion is obtained by replacing the Z$_1$ and Y$_1$ bonds by the Y$_1$ and X$_1$ bonds, respectively. The motion after the next is obtained by replacing the Y$_1$ and X$_1$ bonds by the X$_1$ and Z$_1$ bonds, respectively. On average, a hole in the $J_{\mathrm{eff}}=\frac32$ subspace hops on the X$_1$ bond, on the Y$_1$ bond, and then on the Z$_1$ bond as time advances [Fig.~\ref{fig:honeycomb_hoppings}(b)]. Thus, these two types of spin-orbit excitons have different sequences of the types of bonds on which a hole hops. This implies that the order of the hopping matrices in Eq.~(\ref{eq:expan2approx}) is reversed for a double spin-orbit exciton. Consequently, the effective magnetic field is reversed when double spin-orbit excitons are produced. This would be the reason why the $\perp$ component is reversed for $\omega$=0.7. 

To judge how efficiently the pseudospins are directed to the $\perp$ direction, we plot the averaged magnitude of the pseudospin densities $\frac12 (\mid \mbox{\boldmath $ j $}^{(1/2)}_{\mathrm{site A}} \mid+\mid \mbox{\boldmath $ j $}^{(1/2)}_{\mathrm{site B}} \mid)$ in Fig.~\ref{fig:strong_f_ps_w_dep}(b). From the comparison with Fig.~\ref{fig:strong_f_ps_w_dep}(a), the $\perp$ component is close to the magnitude during the photoexcitation for $\omega$=0.3, indicating that the pseudospins at sites A and B are soon directed nearly to the $\perp$ direction. This is in contrast to the case of weak light fields. 
To observe how the pseudospins are aligned, we plot the cosine of the angle between the pseudospin densities at sites A and B $\mbox{\boldmath $ j $}^{(1/2)}_{\mathrm{site A}} \cdot \mbox{\boldmath $ j $}^{(1/2)}_{\mathrm{site B}}/\left(\mid \mbox{\boldmath $ j $}^{(1/2)}_{\mathrm{site A}} \mid \cdot \mid \mbox{\boldmath $ j $}^{(1/2)}_{\mathrm{site B}} \mid \right)$ in Fig.~\ref{fig:strong_f_ps_w_dep}(c). Note that the cosine is $-1$ when they are antiparallel, and it is 1 when they are parallel. The time profile of the angle sensitively depends on $\omega$. For $\omega$=0.3, the configuration of the pseudospins is changed from antiparallel at $t$=0 to parallel, although their angle occasionally becomes wide at an early stage. For $\omega$=0.5, the configuration of the pseudospins oscillates between nearly antiparallel and nearly parallel. For $\omega$=0.7, where the averaged $\perp$ component acquires a negative value [Fig.~\ref{fig:strong_f_ps_w_dep}(a)], the configuration of the pseudospins is rapidly changed from antiparallel at $t$=0 to nearly parallel at an early stage, although the configuration fluctuates on a short time scale. The increment in the total energy is fast for $\omega$=0.7, as shown later, and the system seems rapidly thermalized by photoexcitation. For $\omega$=0.9, the pseudospin configuration is more complicated and seems rather chaotic. 

\begin{figure}
\includegraphics[height=20.4cm]{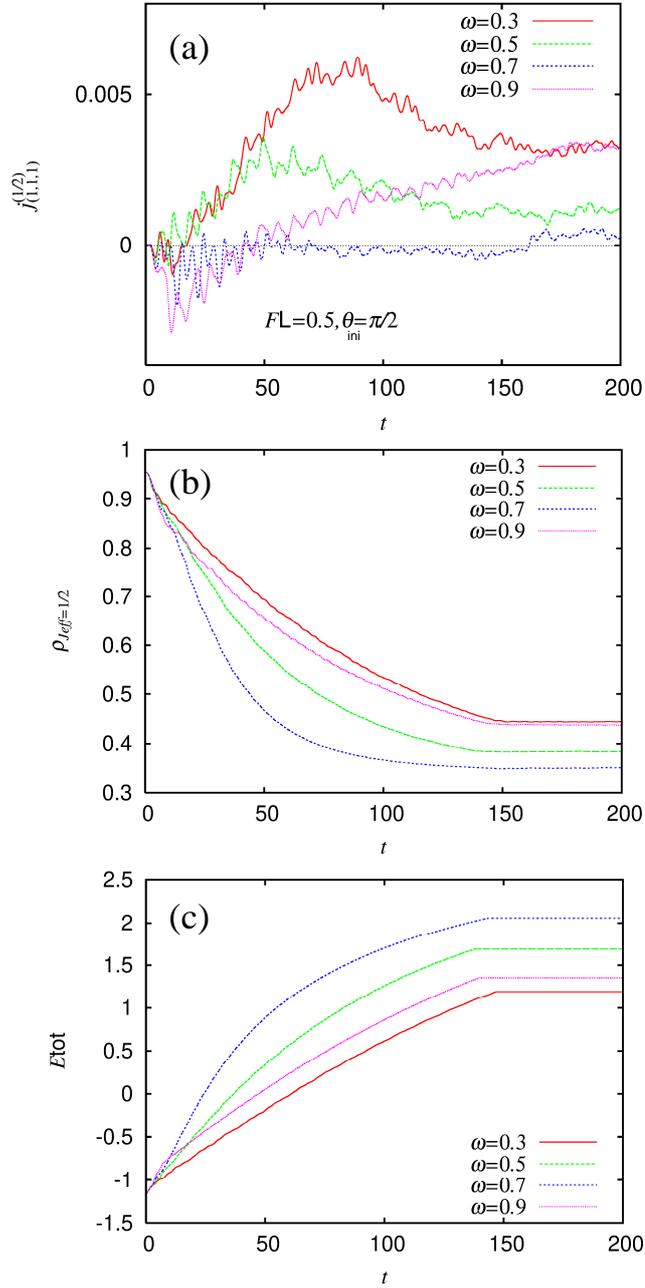}
\caption{(Color online) 
Time profiles of (a) averaged $\perp$ component of $J_{\mathrm{eff}}=\frac12$ pseudospins, (b) averaged charge density in $J_{\mathrm{eff}}=\frac12$ subspace, and (c) total energy, when excited by a circularly polarized light field with $F_{\mathrm{L}}$=0.5 and $\theta_\text{ini}$=$\pi/2$. The duration of photoexcitation is the same as that in Fig.~\ref{fig:strong_f_ps_w_dep}. 
\label{fig:very_strong_f_ps_w_dep}}
\end{figure}
To clarify the relationship with the amount of energy supplied by the light field, the time profiles are plotted for a stronger light field in Fig.~\ref{fig:very_strong_f_ps_w_dep}. For $\omega$=0.3, the averaged $\perp$ component of the pseudospins $\frac12 (j^{(1/2)}_{\mathrm{site A},\perp}+j^{(1/2)}_{\mathrm{site B},\perp})$ reaches a maximum value and then decreases before the light field is switched off [Fig.~\ref{fig:very_strong_f_ps_w_dep}(a)]. The averaged $\perp$ component of the $J_{\mathrm{eff}}=\frac32$ pseudospins $\frac12 (j^{(3/2)}_{\mathrm{site A},\perp}+j^{(3/2)}_{\mathrm{site B},\perp})$ behaves in a similar manner (not shown), and its maximum value ($\simeq$0.011) is about twice as large as that of the $J_{\mathrm{eff}}=\frac12$ pseudospins shown here. The time profiles of the averaged charge density in the $J_{\mathrm{eff}}=\frac12$ subspace $\frac12 (\rho^{(1/2)}_{\mathrm{site A}}+\rho^{(1/2)}_{\mathrm{site B}}) $ and the total energy $\langle H \rangle$ are shown in Figs.~\ref{fig:very_strong_f_ps_w_dep}(b) and \ref{fig:very_strong_f_ps_w_dep}(c), respectively. 
For  $\omega$=0.3, at the time when the averaged $\perp$ component of the pseudospins reaches a maximum value, holes have substantially been transferred to the $J_{\mathrm{eff}}=\frac32$ subspace. Further application of a light field transfers more holes to the $J_{\mathrm{eff}}=\frac32$ subspace, but it produces the opposite effect on the $\perp$ component. This finding suggests that spin-orbit excitons are directly involved with the growth and decay of the $\perp$ component. 
Generally, when the decrease in $\rho^{(1/2)}_{i}$ is rapid, the increase in $\langle H \rangle$ is also rapid. This correlation between the decreasing rate of $\rho^{(1/2)}_{i}$ and the increasing rate of $\langle H \rangle$ is observed for a broad range of $F_{\mathrm{L}}$ and $\omega$ below the Mott gap. 

The $\omega$-dependent dynamics of the $\perp$ component originates from the fact that different types of spin-orbit excitons are produced. When $\omega$ is increased from 0.3 through 0.5 to 0.7, the rate of charge transfer to the $J_{\mathrm{eff}}=\frac32$ subspace is increased. In particular, for $\omega$=0.7, the rate of charge transfer is significantly larger than in the other cases. This is consistent with the picture that double spin-orbit excitons are produced for $\omega$=0.7, which has been presented along with Fig.~\ref{fig:honeycomb_hoppings}. When $\omega$ is further increased to 0.9, the rate becomes small. When $\omega$ is increased from 1.0 through 1.1 to 1.2 above the Mott gap, the initial change rate of the averaged $\perp$ component for $0<t<10$ is almost the same as that for $\omega$=0.9, but the $\perp$ component quickly turns to a sudden decrease and almost vanishes, which appears to be due to thermalization. For the pseudospins, doublons that are produced with $ \omega $ above the Mott gap would act as a heat bath. The timescale of the quick decay process is consistent with this picture (not shown). 

\begin{figure}
\includegraphics[height=13.6cm]{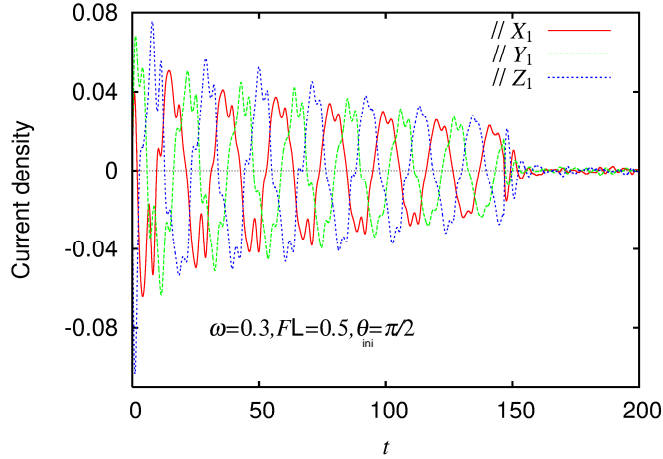}
\caption{(Color online) 
Time profiles of components of current density that are parallel to {\boldmath $X_1$}, {\boldmath $Y_1$}, and {\boldmath $Z_1$} in Fig.~\ref{fig:honeycomb}, when excited by a circularly polarized light field with $\omega$=0.3, $F_{\mathrm{L}}$=0.5, and $\theta_\text{ini}$=$\pi/2$. The duration of photoexcitation is $0<t<t_{\mathrm{off}}\simeq$147 ($n$=7). 
\label{fig:rucl3_circ_u3p0ud1p8j0p6_w0p3_0p5_00_th0p5piTcur_xyz1}}
\end{figure}
For the same light field as that of $\omega$=0.3 in Fig.~\ref{fig:very_strong_f_ps_w_dep}, the time profiles of the components of the current density $ \mbox{\boldmath $j$} = -\partial \langle H \rangle/\partial \mbox{\boldmath $A$} $ are shown in Fig.~\ref{fig:rucl3_circ_u3p0ud1p8j0p6_w0p3_0p5_00_th0p5piTcur_xyz1}, where the component is parallel to one of {\boldmath $X_1$}, {\boldmath $Y_1$}, and {\boldmath $Z_1$} in Fig.~\ref{fig:honeycomb}. It is shown that the current basically flows in the direction of the electric field at the time: the current flows almost parallel to {\boldmath $X_1$}, {\boldmath $Y_1$}, and {\boldmath $Z_1$} when the electric field is parallel to {\boldmath $X_1$}, {\boldmath $Y_1$}, and {\boldmath $Z_1$}, respectively. Although some fine structures appear for strong light fields, this almost parallel behavior is independent of $\omega$. When we increase $\omega$, the angular change rate of the direction of the current density is accordingly increased. Even when $\omega$ is increased to such an extent that the $\perp$ components of the pseudospins obtain negative values, the direction of the current density is still rotated counterclockwise, following the electric field, as in Fig.~\ref{fig:rucl3_circ_u3p0ud1p8j0p6_w0p3_0p5_00_th0p5piTcur_xyz1}. This characteristic indicates that whether the $\perp$ component has a positive value or a negative value is not simply determined by the rotational direction of the current density. 

\section{Conclusions and Discussion}
Photoinduced dynamics are theoretically investigated in a Hubbard model for $\alpha$-RuCl$_3$, which shows a quantum-spin-liquid state. In view of the fact that light fields are mainly coupled to the charge degrees of freedom, we consider three $t_{2g}$ orbitals per Ru site in the model and employ the parameters in Ref.~\citen{winter_prb16}. The exact diagonalization method is used for the ground state of a three-unit-cell system with threefold-symmetric and periodic boundary conditions. Photoinduced dynamics are obtained by numerically solving the time-dependent Schr\"odinger equation. 

Motivated by a recent observation of the photoinduced inverse Faraday effect,\cite{amano_prr22} we treat circularly polarized light fields. To interpret the photoinduced dynamics, we use a high-frequency expansion in the framework of quantum Floquet theory to evaluate effective magnetic fields. We derive two types of effective magnetic fields, although only optical electric fields are introduced through the Peierls phase. One is produced with the help of spin-orbit coupling and originates from $ H_{\mathrm{F},\mathrm{SO}}^{(3)} $. The other is produced from the commutators among the kinetic operators on the three bonds and originates from $ H_{\mathrm{F}}^{(2)} $. It is independent of spin-orbit coupling. These effective fields are useful for interpreting the pseudospin dynamics under different strengths of the light field. 

The effective magnetic fields $ \mbox{\boldmath $B$}_{\mathrm{eff}}(\mbox{\boldmath $k$}) $ originating from $ H_{\mathrm{F},\mathrm{SO}}^{(3)} $ are odd functions of the momentum of a hole; thus, they are produced by the optical electric field that breaks the inversion symmetry in the momentum distribution of holes. Their behaviors are simplified if we assume the threefold symmetry in the transfer integrals. When the direction of bias in the momentum distribution changes from {\boldmath $X_2$} through {\boldmath $Y_2$} to {\boldmath $Z_2$}, the $x$ component of $ \mbox{\boldmath $B$}_{\mathrm{eff}}(\mbox{\boldmath $k$}) $ is first enhanced, then the $y$ component is similarly enhanced, and finally the $z$ component is similarly enhanced. The momentum dependence of $ \mbox{\boldmath $B$}_{\mathrm{eff}}(\mbox{\boldmath $k$}) $ at site B is obtained by reversing that at site A; thus, these fields at sites A and B are almost antiparallel. When a weak circularly polarized light field is applied, the numerically obtained dynamics of the pseudospins at sites A and B are explained by these behaviors of $ \mbox{\boldmath $B$}_{\mathrm{eff}}(\mbox{\boldmath $k$}) $ at respective sites. 

The effective magnetic field originating from $ H_{\mathrm{F}}^{(2)} $ produces nonzero $\perp$ components of the pseudospins for a circularly polarized light field. They are responsible for the observed inverse Faraday effect.\cite{amano_prr22} The sign of the $\perp$ component is consistent with the effective magnetic field for small $\omega$, but the sign depends on the frequency of the light field. The mechanism for this sign change above a threshold located in the range of $0.60<\omega<0.65$ below the Mott gap is the correlated hopping processes for a double spin-orbit exciton, as explained in Fig.~\ref{fig:honeycomb_hoppings}. During the photoexcitation of left-hand (right-hand) circular polarization, the $\perp$ component grows to a positive (negative) value, at least for $\omega\leq$0.6 below the Mott gap. In the present system of minimum size, however, it is not certain that different types of spin-orbit excitons are quantitatively reproduced in the spectra. This sign change for $\omega$ above a threshold is not experimentally observed.\cite{amano_prr22} In $\alpha$-RuCl$_3$, the hopping parameters for the second-nearest neighbors and the third-nearest neighbors are not so small.\cite{winter_prb16} As a consequence, the correlated hopping processes would not be realized in as pure a form as described in Fig.~\ref{fig:honeycomb_hoppings}. The construction of and calculations in an effective model for restricted degrees of freedom and of larger system sizes would be future important topics. 

\begin{acknowledgments}
The author is grateful to S. Iwai and N. Arakawa for various discussions. 
This work was supported by JSPS KAKENHI Grant No. JP16K05459,                          MEXT Q-LEAP Grant No. JPMXS0118067426, and JST CREST Grant No. JPMJCR1901. 
\end{acknowledgments}

\appendix
\section{Effective Fields from $ H_{\mathrm{F},\mathrm{SO}}^{(3)} $ within $J_{\mathrm{eff}}=\frac32$ Subspace \label{sec:J_32}}
Among the terms in $ H_{\mathrm{F},\mathrm{SO}}^{(3)} $, we here show the terms that act on pseudospins within the $J_{\mathrm{eff}}=\frac32$ subspace, which are denoted by $ H_{\mathrm{F},\mathrm{SO}}^{(3,J_{\mathrm{eff}}=\frac32)} $. This operator consists of $ q_{i,J_z}^\dagger q_{i,J^{\,\prime}_z} $ terms with $ i $=$ j $ or next-nearest-neighbor sites $ i $ and $ j $. Thus, it is the sum of $ H_{\mathrm{F},\mathrm{SO}}^{(3,J_{\mathrm{eff}}=\frac32,\mathrm{A})} $, where sites $ i $ and $ j $ belong to sublattice A, and $ H_{\mathrm{F},\mathrm{SO}}^{(3,J_{\mathrm{eff}}=\frac32,\mathrm{B})} $, where sites $ i $ and $ j $ belong to sublattice B. In momentum space, the operator $ H_{\mathrm{F},\mathrm{SO}}^{(3,J_{\mathrm{eff}}=\frac32,\mathrm{A})} $ is written as 
\begin{equation}
H_{\mathrm{F},\mathrm{SO}}^{(3,J_{\mathrm{eff}}=\frac32,\mathrm{A})} = -
\sum_{\mbox{\boldmath $k$}} 
\left( 
q_{\mathrm{A},\mbox{\boldmath $k$},3/2}^\dagger 
q_{\mathrm{A},\mbox{\boldmath $k$},1/2}^\dagger 
q_{\mathrm{A},\mbox{\boldmath $k$},-1/2}^\dagger 
q_{\mathrm{A},\mbox{\boldmath $k$},-3/2}^\dagger 
\right) \mathbf{M}_{\mathrm{A}} (\mbox{\boldmath $k$}) 
\begin{pmatrix}
q_{\mathrm{A},\mbox{\boldmath $k$},3/2} \\
q_{\mathrm{A},\mbox{\boldmath $k$},1/2} \\
q_{\mathrm{A},\mbox{\boldmath $k$},-1/2} \\
q_{\mathrm{A},\mbox{\boldmath $k$},-3/2} \\
\end{pmatrix}
\;, \label{eq:zeeman_32}
\end{equation}
where the creation and annihilation operators act on sublattice A, and the matrix $ \mathbf{M}_{\mathrm{A}} (\mbox{\boldmath $k$}) $ is given by
\begin{equation}
\mathbf{M}_{\mathrm{A}} (\mbox{\boldmath $k$}) = 
\begin{pmatrix}
C_{\mathrm{out}}(\mbox{\boldmath $k$})
+\frac32 B^z_{\mathrm{out}}(\mbox{\boldmath $k$}) &
\frac{\sqrt{3}}{2} \left( 
B^x_{\mathrm{out}}(\mbox{\boldmath $k$}) 
-i B^y_{\mathrm{out}}(\mbox{\boldmath $k$})  \right) & \ast & \ast \\
\frac{\sqrt{3}}{2} \left( 
B^{x\ast}_{\mathrm{out}}(\mbox{\boldmath $k$}) 
+i B^{y\ast}_{\mathrm{out}}(\mbox{\boldmath $k$})  \right) &
C_{\mathrm{in}}(\mbox{\boldmath $k$})
+\frac12 B^z_{\mathrm{in}}(\mbox{\boldmath $k$}) & 
B^x_{\mathrm{in}}(\mbox{\boldmath $k$}) 
-i B^y_{\mathrm{in}}(\mbox{\boldmath $k$}) & \ast \\
\ast & B^x_{\mathrm{in}}(\mbox{\boldmath $k$}) 
+i B^y_{\mathrm{in}}(\mbox{\boldmath $k$}) &
C_{\mathrm{in}}(\mbox{\boldmath $k$})
-\frac12 B^z_{\mathrm{in}}(\mbox{\boldmath $k$}) &
\frac{\sqrt{3}}{2} \left( 
B^{x\ast}_{\mathrm{out}}(\mbox{\boldmath $k$}) 
-i B^{y\ast}_{\mathrm{out}}(\mbox{\boldmath $k$})  \right) \\
\ast & \ast & \frac{\sqrt{3}}{2} \left( 
B^x_{\mathrm{out}}(\mbox{\boldmath $k$}) 
+i B^y_{\mathrm{out}}(\mbox{\boldmath $k$})  \right) &
C_{\mathrm{out}}(\mbox{\boldmath $k$})
-\frac32 B^z_{\mathrm{out}}(\mbox{\boldmath $k$}) \\
\end{pmatrix}
\;, \label{eq:matrix_32}
\end{equation}
where $ C_{\mathrm{in}}(\mbox{\boldmath $k$}) $ and $ C_{\mathrm{out}}(\mbox{\boldmath $k$}) $ are even functions of {\boldmath $k$}. 
Note that the effective magnetic fields $ B^{x,y,z}_{\mathrm{in,out}}(\mbox{\boldmath $k$}) $ originating from $ H_{\mathrm{F},\mathrm{SO}}^{(3)} $ include the Land\'e $g$-factor for $J_{\mathrm{eff}}=\frac32$. 
Although they are not explicitly shown, the matrix elements denoted by $ \ast $ above are nonzero. The effective magnetic fields appearing above are given by 
\begin{equation}
B^{x,y,z}_{\mathrm{in,out}}(\mbox{\boldmath $k$}) = 
\sum_{m \neq 0} \frac{-\lambda}{2(m\hbar\omega)^2} \times 
\left[ 
X_{m,\mathrm{in,out}} (\mbox{\boldmath $k$}), 
Y_{m,\mathrm{in,out}} (\mbox{\boldmath $k$}), 
Z_{m,\mathrm{in,out}} (\mbox{\boldmath $k$}) \right]
\;, \label{eq:mag_field_32}
\end{equation}
with 
\begin{eqnarray}
X_{m,\mathrm{in}} (\mbox{\boldmath $k$}) & = &
\left( -ie^{i\mbox{\boldmath $k$}\cdot\mbox{\boldmath $Y_2$}}
+ie^{-i\mbox{\boldmath $k$}\cdot\mbox{\boldmath $Y _2$}} \right) 
\left( J_{-m}(-\mbox{\boldmath $Z_1$}) J_{m}(\mbox{\boldmath $X_1$}) 
+ J_{m}(-\mbox{\boldmath $Z_1$}) J_{-m}(\mbox{\boldmath $X_1$}) 
\right) \nonumber \\
&& \times \frac14 \left( 
2t_1 t^{\,\prime}_{2 } +2t_4 t^{\,\prime}_{1b} -2t_3 t^{\,\prime}_{2 }
-t_4 t^{\,\prime}_{1a} -t_4 t^{\,\prime}_{3 }
\right) \nonumber \\
&& + \left( -ie^{i\mbox{\boldmath $k$}\cdot\mbox{\boldmath $Z_2$}}
+ie^{-i\mbox{\boldmath $k$}\cdot\mbox{\boldmath $Z_2$}} \right) 
\left( J_{-m}(-\mbox{\boldmath $X_1$}) J_{m}(\mbox{\boldmath $Y_1$}) 
+ J_{m}(-\mbox{\boldmath $X_1$}) J_{-m}(\mbox{\boldmath $Y_1$}) 
\right) \nonumber \\
&& \times \frac14 \left( 
2t^{\,\prime}_{1b} t^{\,\prime}_{2 } -2t^{\,\prime}_{1b} t^{\,\prime}_{4b}
+t^{\,\prime}_{1a} t^{\,\prime}_{4b}  +t^{\,\prime}_{3 } t^{\,\prime}_{4b}
-t^{\,\prime}_{1a} t^{\,\prime}_{2 }  -t^{\,\prime}_{2 } t^{\,\prime}_{3 }
\right) \nonumber \\
&& + \left( -ie^{i\mbox{\boldmath $k$}\cdot\mbox{\boldmath $X_2$}}
+ie^{-i\mbox{\boldmath $k$}\cdot\mbox{\boldmath $X_2$}} \right) 
\left( J_{-m}(-\mbox{\boldmath $Y_1$}) J_{m}(\mbox{\boldmath $Z_1$}) 
+ J_{m}(-\mbox{\boldmath $Y_1$}) J_{-m}(\mbox{\boldmath $Z_1$}) 
\right) \nonumber \\
&& \times \frac14 \left( 
2t_3 t^{\,\prime}_{4b} -2t_1 t^{\,\prime}_{4b} -2t_4 t^{\,\prime}_{1b}
+t_4 t^{\,\prime}_{1a}  +t_4 t^{\,\prime}_{3 }
\right) 
\;, \label{eq:mag_f_x_in}
\end{eqnarray}
\begin{eqnarray}
Y_{m,\mathrm{in}} (\mbox{\boldmath $k$}) & = &
\left( -ie^{i\mbox{\boldmath $k$}\cdot\mbox{\boldmath $Z_2$}}
+ie^{-i\mbox{\boldmath $k$}\cdot\mbox{\boldmath $Z_2$}} \right) 
\left( J_{-m}(-\mbox{\boldmath $X_1$}) J_{m}(\mbox{\boldmath $Y_1$}) 
+ J_{m}(-\mbox{\boldmath $X_1$}) J_{-m}(\mbox{\boldmath $Y_1$}) 
\right) \nonumber \\
&& \times \frac14 \left( 
2t^{\,\prime}_{1b} t^{\,\prime}_{2 } -2t^{\,\prime}_{1b} t^{\,\prime}_{4b}
+t^{\,\prime}_{1a} t^{\,\prime}_{4b}  +t^{\,\prime}_{3 } t^{\,\prime}_{4b}
-t^{\,\prime}_{1a} t^{\,\prime}_{2 }  -t^{\,\prime}_{2 } t^{\,\prime}_{3 }
\right) \nonumber \\
&& + \left( -ie^{i\mbox{\boldmath $k$}\cdot\mbox{\boldmath $X_2$}}
+ie^{-i\mbox{\boldmath $k$}\cdot\mbox{\boldmath $X_2$}} \right) 
\left( J_{-m}(-\mbox{\boldmath $Y_1$}) J_{m}(\mbox{\boldmath $Z_1$}) 
+ J_{m}(-\mbox{\boldmath $Y_1$}) J_{-m}(\mbox{\boldmath $Z_1$}) 
\right) \nonumber \\
&& \times \frac14 \left( 
2t_1 t^{\,\prime}_{2 } +2t_4 t^{\,\prime}_{1b} -2t_3 t^{\,\prime}_{2 }
-t_4 t^{\,\prime}_{1a} -t_4 t^{\,\prime}_{3 }
\right) \nonumber \\
&& + \left( -ie^{i\mbox{\boldmath $k$}\cdot\mbox{\boldmath $Y_2$}}
+ie^{-i\mbox{\boldmath $k$}\cdot\mbox{\boldmath $Y _2$}} \right) 
\left( J_{-m}(-\mbox{\boldmath $Z_1$}) J_{m}(\mbox{\boldmath $X_1$}) 
+ J_{m}(-\mbox{\boldmath $Z_1$}) J_{-m}(\mbox{\boldmath $X_1$}) 
\right) \nonumber \\
&& \times \frac14 \left( 
2t_3 t^{\,\prime}_{4b} -2t_1 t^{\,\prime}_{4b} -2t_4 t^{\,\prime}_{1b}
+t_4 t^{\,\prime}_{1a}  +t_4 t^{\,\prime}_{3 }
\right) 
\;, \label{eq:mag_f_y_in}
\end{eqnarray}
\begin{eqnarray}
Z_{m,\mathrm{in}} (\mbox{\boldmath $k$}) & = & 
\left[ \left( -ie^{i\mbox{\boldmath $k$}\cdot\mbox{\boldmath $X_2$}}
+ie^{-i\mbox{\boldmath $k$}\cdot\mbox{\boldmath $X_2$}} \right) 
\left( J_{-m}(-\mbox{\boldmath $Y_1$}) J_{m}(\mbox{\boldmath $Z_1$}) 
+ J_{m}(-\mbox{\boldmath $Y_1$}) J_{-m}(\mbox{\boldmath $Z_1$}) \right) 
\right. \nonumber \\
&& \left. +\left( -ie^{i\mbox{\boldmath $k$}\cdot\mbox{\boldmath $Y_2$}}
+ie^{-i\mbox{\boldmath $k$}\cdot\mbox{\boldmath $Y_2$}} \right) 
\left( J_{-m}(-\mbox{\boldmath $Z_1$}) J_{m}(\mbox{\boldmath $X_1$}) 
+ J_{m}(-\mbox{\boldmath $Z_1$}) J_{-m}(\mbox{\boldmath $X_1$}) \right) 
\right] \nonumber \\
&& \times \frac32 \left( 
t_4 t^{\,\prime}_{2 } -t_4 t^{\,\prime}_{4b} 
\right) \nonumber \\
&& +\left( -ie^{i\mbox{\boldmath $k$}\cdot\mbox{\boldmath $Z_2$}}
+ie^{-i\mbox{\boldmath $k$}\cdot\mbox{\boldmath $Z_2$}} \right) 
\left( J_{-m}(-\mbox{\boldmath $X_1$}) J_{m}(\mbox{\boldmath $Y_1$}) 
+ J_{m}(-\mbox{\boldmath $X_1$}) J_{-m}(\mbox{\boldmath $Y_1$}) 
\right) \nonumber \\
&& \times \frac32 \left( 
4t^{\,\prime}_{4b} t^{\,\prime}_{4b} -4t^{\,\prime}_{2 } t^{\,\prime}_{2 }
\right)
\;, \label{eq:mag_f_z_in}
\end{eqnarray}
\begin{eqnarray}
X_{m,\mathrm{out}} (\mbox{\boldmath $k$}) & = &
-ie^{i\mbox{\boldmath $k$}\cdot\mbox{\boldmath $Y_2$}}
\left( J_{-m}(-\mbox{\boldmath $Z_1$}) J_{m}(\mbox{\boldmath $X_1$}) 
+ J_{m}(-\mbox{\boldmath $Z_1$}) J_{-m}(\mbox{\boldmath $X_1$}) 
\right) \nonumber \\
&& \times \frac16 \left( 
6t_2 t^{\,\prime}_{4b} +2t_4 t^{\,\prime}_{1b}
-t_4 t^{\,\prime}_{1a}  -t_4 t^{\,\prime}_{3 }
\right) \nonumber \\
&& +ie^{-i\mbox{\boldmath $k$}\cdot\mbox{\boldmath $Y _2$}}
\left( J_{-m}(-\mbox{\boldmath $Z_1$}) J_{m}(\mbox{\boldmath $X_1$}) 
+ J_{m}(-\mbox{\boldmath $Z_1$}) J_{-m}(\mbox{\boldmath $X_1$}) 
\right) \nonumber \\
&& \times \frac16 \left( 
-6t_4 t^{\,\prime}_{4a} +3t_4 t^{\,\prime}_{3 } -3t_4 t^{\,\prime}_{1a}
+2t_1 t^{\,\prime}_{2 } -2t_3 t^{\,\prime}_{2 }
\right) \nonumber \\
&& -ie^{i\mbox{\boldmath $k$}\cdot\mbox{\boldmath $Z_2$}}
\left( J_{-m}(-\mbox{\boldmath $X_1$}) J_{m}(\mbox{\boldmath $Y_1$}) 
+ J_{m}(-\mbox{\boldmath $X_1$}) J_{-m}(\mbox{\boldmath $Y_1$}) 
\right) \nonumber \\
&& \times \frac16 \left( 
 6t^{\,\prime}_{2 } t^{\,\prime}_{4a} +3t^{\,\prime}_{1a} t^{\,\prime}_{4b}
-3t^{\,\prime}_{3 } t^{\,\prime}_{4b} +2t^{\,\prime}_{1b} t^{\,\prime}_{2 }
 -t^{\,\prime}_{1a} t^{\,\prime}_{2 }  -t^{\,\prime}_{2 } t^{\,\prime}_{3 }
\right) \nonumber \\
&& +ie^{-i\mbox{\boldmath $k$}\cdot\mbox{\boldmath $Z_2$}} 
\left( J_{-m}(-\mbox{\boldmath $X_1$}) J_{m}(\mbox{\boldmath $Y_1$}) 
+ J_{m}(-\mbox{\boldmath $X_1$}) J_{-m}(\mbox{\boldmath $Y_1$}) 
\right) \nonumber \\
&& \times \frac16 \left( 
-6t^{\,\prime}_{4a} t^{\,\prime}_{4b} +3t^{\,\prime}_{1a} t^{\,\prime}_{2 }
-3t^{\,\prime}_{2 } t^{\,\prime}_{3 } -2t^{\,\prime}_{1b} t^{\,\prime}_{4b}
 +t^{\,\prime}_{1a} t^{\,\prime}_{4b}  +t^{\,\prime}_{3 } t^{\,\prime}_{4b}
\right) \nonumber \\
&& -ie^{i\mbox{\boldmath $k$}\cdot\mbox{\boldmath $X_2$}}
\left( J_{-m}(-\mbox{\boldmath $Y_1$}) J_{m}(\mbox{\boldmath $Z_1$}) 
+ J_{m}(-\mbox{\boldmath $Y_1$}) J_{-m}(\mbox{\boldmath $Z_1$}) 
\right) \nonumber \\
&& \times \frac16 \left( 
 6t_4 t^{\,\prime}_{4a} +3t_4 t^{\,\prime}_{3 } -3t_4 t^{\,\prime}_{1a}
+2t_3 t^{\,\prime}_{4b} -2t_1 t^{\,\prime}_{4b}
\right) \nonumber \\
&& +ie^{-i\mbox{\boldmath $k$}\cdot\mbox{\boldmath $X_2$}} 
\left( J_{-m}(-\mbox{\boldmath $Y_1$}) J_{m}(\mbox{\boldmath $Z_1$}) 
+ J_{m}(-\mbox{\boldmath $Y_1$}) J_{-m}(\mbox{\boldmath $Z_1$}) 
\right) \nonumber \\
&& \times \frac16 \left( 
-6t_2 t^{\,\prime}_{2 } -2t_4 t^{\,\prime}_{1b}
+ t_4 t^{\,\prime}_{1a} + t_4 t^{\,\prime}_{3 }
\right) 
\;, \label{eq:mag_f_x_out}
\end{eqnarray}
\begin{eqnarray}
Y_{m,\mathrm{out}} (\mbox{\boldmath $k$}) & = &
-ie^{i\mbox{\boldmath $k$}\cdot\mbox{\boldmath $Z_2$}}
\left( J_{-m}(-\mbox{\boldmath $X_1$}) J_{m}(\mbox{\boldmath $Y_1$}) 
+ J_{m}(-\mbox{\boldmath $X_1$}) J_{-m}(\mbox{\boldmath $Y_1$}) 
\right) \nonumber \\
&& \times \frac16 \left( 
-6t^{\,\prime}_{4a} t^{\,\prime}_{4b} +3t^{\,\prime}_{1a} t^{\,\prime}_{2 }
-3t^{\,\prime}_{2 } t^{\,\prime}_{3 } -2t^{\,\prime}_{1b} t^{\,\prime}_{4b}
 +t^{\,\prime}_{1a} t^{\,\prime}_{4b}  +t^{\,\prime}_{3 } t^{\,\prime}_{4b}
\right) \nonumber \\
&& +ie^{-i\mbox{\boldmath $k$}\cdot\mbox{\boldmath $Z_2$}}
\left( J_{-m}(-\mbox{\boldmath $X_1$}) J_{m}(\mbox{\boldmath $Y_1$}) 
+ J_{m}(-\mbox{\boldmath $X_1$}) J_{-m}(\mbox{\boldmath $Y_1$}) 
\right) \nonumber \\
&& \times \frac16 \left( 
6t^{\,\prime}_{2 } t^{\,\prime}_{4a} +3t^{\,\prime}_{1a} t^{\,\prime}_{4b}
-3t^{\,\prime}_{3 } t^{\,\prime}_{4b} +2t^{\,\prime}_{1b} t^{\,\prime}_{2 }
 -t^{\,\prime}_{1a} t^{\,\prime}_{2 }  -t^{\,\prime}_{2 } t^{\,\prime}_{3 }
\right) \nonumber \\
&& -ie^{i\mbox{\boldmath $k$}\cdot\mbox{\boldmath $X_2$}}
\left( J_{-m}(-\mbox{\boldmath $Y_1$}) J_{m}(\mbox{\boldmath $Z_1$}) 
+ J_{m}(-\mbox{\boldmath $Y_1$}) J_{-m}(\mbox{\boldmath $Z_1$}) 
\right) \nonumber \\
&& \times \frac16 \left( 
-6t_4 t^{\,\prime}_{4a} +3t_4 t^{\,\prime}_{3 } -3t_4 t^{\,\prime}_{1a}
+2t_1 t^{\,\prime}_{2 } -2t_3 t^{\,\prime}_{2 }
\right) \nonumber \\
&& +ie^{-i\mbox{\boldmath $k$}\cdot\mbox{\boldmath $X_2$}} 
\left( J_{-m}(-\mbox{\boldmath $Y_1$}) J_{m}(\mbox{\boldmath $Z_1$}) 
+ J_{m}(-\mbox{\boldmath $Y_1$}) J_{-m}(\mbox{\boldmath $Z_1$}) 
\right) \nonumber \\
&& \times \frac16 \left( 
6t_2 t^{\,\prime}_{4b} +2t_4 t^{\,\prime}_{1b}
-t_4 t^{\,\prime}_{1a}  -t_4 t^{\,\prime}_{3 }
\right) \nonumber \\
&& -ie^{i\mbox{\boldmath $k$}\cdot\mbox{\boldmath $Y_2$}}
\left( J_{-m}(-\mbox{\boldmath $Z_1$}) J_{m}(\mbox{\boldmath $X_1$}) 
+ J_{m}(-\mbox{\boldmath $Z_1$}) J_{-m}(\mbox{\boldmath $X_1$}) 
\right) \nonumber \\
&& \times \frac16 \left( 
-6t_2 t^{\,\prime}_{2 } -2t_4 t^{\,\prime}_{1b}
+ t_4 t^{\,\prime}_{1a} + t_4 t^{\,\prime}_{3 }
\right) \nonumber \\
&& +ie^{-i\mbox{\boldmath $k$}\cdot\mbox{\boldmath $Y_2$}} 
\left( J_{-m}(-\mbox{\boldmath $Z_1$}) J_{m}(\mbox{\boldmath $X_1$}) 
+ J_{m}(-\mbox{\boldmath $Z_1$}) J_{-m}(\mbox{\boldmath $X_1$}) 
\right) \nonumber \\
&& \times \frac16 \left( 
6t_4 t^{\,\prime}_{4a} +3t_4 t^{\,\prime}_{3 } -3t_4 t^{\,\prime}_{1a}
+2t_3 t^{\,\prime}_{4b} -2t_1 t^{\,\prime}_{4b}
\right) 
\;, \label{eq:mag_f_y_out}
\end{eqnarray}
\begin{eqnarray}
Z_{m,\mathrm{out}} (\mbox{\boldmath $k$}) & = & 
\left[ \left( -ie^{i\mbox{\boldmath $k$}\cdot\mbox{\boldmath $X_2$}}
+ie^{-i\mbox{\boldmath $k$}\cdot\mbox{\boldmath $X_2$}} \right) 
\left( J_{-m}(-\mbox{\boldmath $Y_1$}) J_{m}(\mbox{\boldmath $Z_1$}) 
+ J_{m}(-\mbox{\boldmath $Y_1$}) J_{-m}(\mbox{\boldmath $Z_1$}) \right) 
\right. \nonumber \\
&& \left. +\left( -ie^{i\mbox{\boldmath $k$}\cdot\mbox{\boldmath $Y_2$}}
+ie^{-i\mbox{\boldmath $k$}\cdot\mbox{\boldmath $Y_2$}} \right) 
\left( J_{-m}(-\mbox{\boldmath $Z_1$}) J_{m}(\mbox{\boldmath $X_1$}) 
+ J_{m}(-\mbox{\boldmath $Z_1$}) J_{-m}(\mbox{\boldmath $X_1$}) \right) 
\right] \nonumber \\
&& \times \frac16 \left( 
2t_2 t^{\,\prime}_{1a} -2t_2 t^{\,\prime}_{3 } 
+t_4 t^{\,\prime}_{2 }  -t_4 t^{\,\prime}_{4b} 
\right) \nonumber \\
&& +\left( -ie^{i\mbox{\boldmath $k$}\cdot\mbox{\boldmath $Z_2$}}
+ie^{-i\mbox{\boldmath $k$}\cdot\mbox{\boldmath $Z_2$}} \right) 
\left( J_{-m}(-\mbox{\boldmath $X_1$}) J_{m}(\mbox{\boldmath $Y_1$}) 
+ J_{m}(-\mbox{\boldmath $X_1$}) J_{-m}(\mbox{\boldmath $Y_1$}) 
\right) \nonumber \\
&& \times \frac16 \left( 
4t^{\,\prime}_{3 } t^{\,\prime}_{4a} -4t^{\,\prime}_{1a} t^{\,\prime}_{4a}
+t^{\,\prime}_{4b} t^{\,\prime}_{4b}  -t^{\,\prime}_{2 } t^{\,\prime}_{2 }
\right)
\;. \label{eq:mag_f_z_out}
\end{eqnarray}
Since $ B^{x,y}_{\mathrm{out}}(\mbox{\boldmath $k$}) $ in Eq.~(\ref{eq:matrix_32}) are complex quantities, the imaginary (and {\boldmath $k$}-even) part of $ B^{y(x)}_{\mathrm{out}}(\mbox{\boldmath $k$}) $ should be added to or subtracted from the real (and {\boldmath $k$}-odd) part of $ B^{x(y)}_{\mathrm{out}}(\mbox{\boldmath $k$}) $ to be interpreted as the $x$ ($y$) component of a field. Thus, the resultant field appearing in the (1,2) and (2,1) elements of $ \mathbf{M}_{\mathrm{A}} (\mbox{\boldmath $k$}) $ is different from that appearing in the (3,4) and (4,3) elements. 
The operator $ H_{\mathrm{F},\mathrm{SO}}^{(3,J_{\mathrm{eff}}=\frac32,\mathrm{B})} $ is given by $ \mathbf{M}_{\mathrm{A}} (\mbox{\boldmath $k$}) \rightarrow \mathbf{M}_{\mathrm{B}} (\mbox{\boldmath $k$})=\mathbf{M}_{\mathrm{A}} (-\mbox{\boldmath $k$}) $ in Eq.~(\ref{eq:zeeman_32}) and $ q_{\mathrm{A},\mbox{\boldmath $k$},\sigma} \rightarrow q_{\mathrm{B},\mbox{\boldmath $k$},\sigma} $, $ q_{\mathrm{A},\mbox{\boldmath $k$},\sigma}^\dagger \rightarrow q_{\mathrm{B},\mbox{\boldmath $k$},\sigma}^\dagger $ for $ \sigma=3/2,1/2,-1/2,$ and $-3/2 $; the creation and annihilation operators now act on sublattice B. 
Here also, the effective magnetic fields have the properties originating from the symmetry mentioned in Sect.~\ref{sect:reflection}. The simultaneous exchanges of vectors {\boldmath $X_1$} for {\boldmath $Y_1$}, {\boldmath $X_2$} for $ -\mbox{\boldmath $Y_2$} $, and {\boldmath $Z_2$} for $ -\mbox{\boldmath $Z_2$} $ lead to the exchanges of $ B^x_{\mathrm{in}}(\mbox{\boldmath $k$}) $ for $ -B^y_{\mathrm{in}}(\mbox{\boldmath $k$}) $, $ B^z_{\mathrm{in}}(\mbox{\boldmath $k$}) $ for $ -B^z_{\mathrm{in}}(\mbox{\boldmath $k$}) $, $ B^x_{\mathrm{out}}(\mbox{\boldmath $k$}) $ for $ -B^y_{\mathrm{out}}(\mbox{\boldmath $k$}) $, and $ B^z_{\mathrm{out}}(\mbox{\boldmath $k$}) $ for $ -B^z_{\mathrm{out}}(\mbox{\boldmath $k$}) $. 

\bibliography{70606}

\end{document}